\documentclass[5p, authoryear,nopreprintline]{elsarticle} 

\usepackage[hyphens]{url}

\journal{Journal of Systems and Software} 

\usepackage{lineno} 

\usepackage{graphicx}

\usepackage{amsmath}
\usepackage{float}
\usepackage[utf8]{inputenc}
\usepackage{booktabs}
\usepackage{tabularx}
\usepackage{xcolor}

\usepackage{natbib}
\usepackage{hyperref}
\usepackage{tcolorbox}

\begin{document}

\urlstyle{same}

\begin{frontmatter}

  \title{On the Dependency Heaviness of CRAN/Bioconductor Ecosystem}
    \author[1]{Zuguang Gu\corref{cor1}}
   \ead{z.gu@dkfz.de} 
      \affiliation[1]{organization={National Center for Tumor Diseases (NCT)},
            addressline={Im Neuenheimer Feld 460}, 
            city={Heidelberg},
            postcode={69120}, 
            country={Germany}}
    \cortext[cor1]{Corresponding author}
  
  \begin{abstract}
  The R package ecosystem is expanding fast and dependencies among
  packages are becoming more complex in the ecosystem. In this study, we
  explored the package dependencies from a new aspect. We applied a new
  metric named ``dependency heaviness'' which measures the number of
  additional strong dependencies that a package uniquely contributes to
  its child or downstream packages. We systematically studied how the
  dependency heaviness spreads from parent to child packages, and how it
  further spreads to remote downstream packages in the CRAN/Bioconductor
  ecosystem. We extracted top packages and key paths that majorly
  transmit heavy dependencies in the ecosystem. Additionally, the
  dependency heaviness analysis on the ecosystem has been implemented as
  a web-based database that provides comprehensive tools for querying
  dependencies of individual R packages.
  \end{abstract}
  
 \end{frontmatter}

\section{Introduction}\label{introduction}

R has become a major programming language applied in many fields,
including statistics, bioinformatics, geoinformatics, economics and
general data science. It is widely used for a variety of tasks such as
data processing \citep{r4ds}, visualization \citep{fdv}, 
statistical modeling \citep{mass}, interactive web
application development \citep{shiny} and reproducible reporting \citep{knitr}. 
The reusable and extensible code implemented by developers are
formatted as R packages and distributed on public repositories such as
the Comprehensive R Archive Network (CRAN)\footnote{\url{https://cran.r-project.org/}.} and Bioconductor\footnote{\url{http://bioconductor.org/}.} \citep{Huber2015}. 
CRAN is the major repository for R packages and Bioconductor
specifically focuses on biology data analysis, especially on
high-throughput omics data\footnote{Omics is a branch of biology science which studies biological systems from large-scale data. It includes a list of sub-branches, such as genomics which aims to reveal mutations and rearrangements on DNAs, and transcriptomics which quantifies mRNA levels of all genes in an organism.}. Both repositories perform manual validations
on new packages and apply regular checks on packages to ensure their
quality and usability. The number of packages in the two repositories
increases almost exponentially by year \citep{changesCRAN}. Besides these two ``standard repositories'', there are also a
huge number of R packages distributed on GitHub, which are
self-maintained and mainly for experimental purposes.

Code reuse is an essential part of a programming language \citep{Haefliger}. An R package
may inherit functionalities from other packages, which forms the
dependency relations between packages. In other words, a package depends
on the availability of other packages to work. R packages as well as
their complex dependency relations construct the ecosystem. The
continuously increasing number of R packages makes the dependency
relations among packages even more complex \citep{Mora_evolution}. By 2022-06-08, there are in
total 22,076 CRAN and Bioconductor packages, including 194,351 direct
dependency
relations\footnote{The statistics were obtained with the R function \texttt{available.packages()}.}.
With such complex dependency relations, the ecosystem might be
vulnerable to failures of only a small number of packages. For example,
recently a compiling failure of the \emph{isoband} package caused
additional failures of 4,747 (almost 25\%) of all packages on
CRAN\footnote{See a bug report on \url{https://twitter.com/cjvanlissa/status/1577552826561171457}. The failure was fixed very quickly within a week, \url{https://github.com/wilkelab/isoband/issues/33}.}.
\emph{isoband} is a dependency of the hub package \emph{ggplot2} and it results in the
failure being transmitted continuously to a huge number of downstream
packages via \emph{ggplot2}. Thus, it is important to explore the
dependency structure in the R package ecosystem from the aspect of
software engineering, which helps to understand the structure of the
ecosystem, to reveal top packages having major impacts and to study the
vulnerability of the ecosystem.

The ecosystem contains complex dependency relations between packages,
which can be naturally modeled as a graph \citep{Miguel}. There are mainly two types of
studies on the dependency graph. In the first category, researchers
applied complex network analysis (CNA) approaches on the dependency
graph, such as calculating various centrality metrics to explore the
impacts of top packages, or applying network community methods to
partition packages into densely connected subgraphs to further analyze
their specific attributes \citep{Mora_CNA}. In the second category, researchers developed
tools for visualizing the complex dependency relations. These tools are
mainly implemented as R packages, such as \emph{deepdep} \citep{deepdep}, 
\emph{pkgnet} \citep{pkgnet}, \emph{pkggraph}
\citep{pkggraph} and \emph{miniCRAN} \citep{miniCRAN}. They
give intuitive views of how dependencies are transmitted between
packages. Nevertheless, they only work well on subgraphs showing local
relations with small sizes, e.g., dependencies from a small number of R
packages, while it becomes difficult to generate and to read when the
size of the graph increases.

Network analysis on the dependency graph helps to understand the R
ecosystem from the system's level. The degree centrality is a
widely-studied metric which measures the number of dependency packages
or the number of child packages that depend on a package (i.e., the
dependents) \citep{Korkmaz2019, Mora_CNA}. Indeed, degree is an important metric for
revealing top packages that have significant impacts on the ecosystem.
However, it is a local metric and we can still look at the system from
new aspects. In practice, when a user installs a new package \emph{P},
additional packages to be installed that he would notice are actually
the total dependency packages upstream of \emph{P}, while which
packages are \emph{P}'s direct dependencies are unobservable to the
user. From a developer's perspective, the direct dependencies of his
package provide no information of which parent brings more dependencies
to it, while he needs to inspect the complete upstream of the dependency
chain to find out parent packages contributing heavy dependencies. All
these imply, the number of ``total dependencies'' instead of ``direct
dependencies'' is a more practical metric. In fact, the number of total
dependencies begins to be paid more attention in the R community. For
example, Bioconductor and R universe\footnote{\url{https://r-universe.dev/}.}
are listing the number of total dependencies as a basic metric for the packages hosted there.

When a package \emph{P} has potentially more total dependencies, there
are several consequences which affect the usability of \emph{P}. We have
listed the risks in our previous paper \citep{pkgndep}:
\emph{``(i) Users have to install a lot of additional packages when
installing P, which would bring the risk that installation failure of
any upstream package stops the installation of P. (ii) The number of
packages loaded into the R session after loading P will be huge, which
increases the difficulty to reproduce a completely identical working
environment on other computers. (iii) Dependencies of P will spread to
all its child packages. (iv) On the platforms for continuous integration
such as GitHub Action or Travis CI, automatic validation of P could
easily fail due to the failures of its upstream packages.''} Therefore,
it is important to reveal packages contributing high total dependencies
to better study their impacts on the vulnerability of the ecosystem.

Total dependency, or the transitive dependency, has already been
investigated in several studies \citep{abate, Mora_CNA}, however, there still lacks
a way to capture the unique transitive dependencies that a single package
contributes. In our previous study \citep{pkgndep}, we proposed
a new metric named ``\emph{dependency heaviness}'' that measures the
number of dependencies that a parent uniquely brings to its child
package and are not brought by any other parent. Simply speaking, this
new metric helps to identify which parent package is heavier in the
context of how it uniquely contributes the dependencies to its child
package. Since now the dependency contribution of parent packages can be
measured quantitatively, developers can easily identify heavy parents of
their packages, then apply possible optimization to reduce the
complexity of package dependencies and build more robust software. Of
course, how to optimize the dependency depends on the specific uses of
parent packages in the corresponding package. We recommended several
solutions in Section \ref{parent_developer}.

Dependency heaviness can provide new insights for risk analysis on
package ecosystems. Number of dependencies, especially transitive
dependencies, is a commonly-used metric of how vulnerable a package is 
to code breaks in the ecosystem. Then, the dependency heaviness also measures
accumulated risks from upstream to a package that are uniquely
transmitted via a parent.

We have implemented the dependency heaviness metric in an R package
\emph{pkgndep}\footnote{\url{https://CRAN.R-project.org/package=pkgndep}.}. In our previous study \citep{pkgndep}, we
briefly described \emph{pkgndep} as software with several use cases. In
this study, we extended the definition of dependency heaviness to a
broader range. Besides the dependency heaviness from a single parent to
a single child package, we also studied how it is transmitted to remote 
downstream packages. We performed a systematic empirical
study on the dependency transmission in the R package ecosystem. The
contributions of this study are briefly listed as follows:

\begin{enumerate}
\def\labelenumi{\arabic{enumi}.}
\item
  We studied how dependency heaviness spreads locally from parent to
  child packages, and we studied how dependency heaviness is
  simultaneously contributed by two parent packages.
\item
  We studied how dependency heaviness is transmitted remotely from
  upstream to downstream packages.
\item
  We applied CNA approaches and we extracted top packages and key paths
  that majorly transmit heavy dependencies in the ecosystem.
\item
  The dependency heaviness analysis on the ecosystem has been
  implemented as a web-based database which provides comprehensive tools
  for analyzing dependencies of individual R packages.
\end{enumerate}

The remainder of this paper is organized as follows. Section
\ref{current_studies} briefly summarizes current studies on the
ecosystem of R as well as other programming languages. Section
\ref{background} provides technical background for understanding
dependency relations in R. Section \ref{methods} introduces definitions
of various heaviness metrics. Section \ref{data} describes the new
functionalities of the \emph{pkgndep} package for this study as well as
data processing. Section \ref{rq} raises various
research questions. Section \ref{results} describes the results of the analysis and answers the research questions. Section \ref{database} describes the web-based
database for the dependency heaviness analysis. Section
\ref{consideration_for_developers} discusses how dependency
heaviness analysis benefits developers. Section \ref{discussion} summarizes the study. 
Section \ref{future} discusses
limitations of the analysis and proposes future plans. Section
\ref{conclusion} encloses the paper with conclusions.

\section{Current studies}\label{current_studies}

\subsection{Current studies on the R package
ecosystem}\label{current-studies-on-the-r-package-ecosystem}

There are a few studies applied on the R ecosystem where researchers
analyzed the dependency relations as well as from other aspects. In this
section, we briefly described their analyses and findings.

R package dependencies can be modeled as a network with a complex
structure. \citet{Mora_CNA} performed complex network analysis on the CRAN ecosystem. Similar
to social networks, they found the dependency network of CRAN also has
the scale-free property. In a scale-free network, node degree follows a
power-law distribution and the proportion of degrees of hub nodes is
stable as the network size increases \citep{wang}. This implies,
in the R ecosystem, there are only a small number of hub packages that
contribute huge amounts of dependencies to other packages. They also
partitioned the global dependency network into modules and they found
CRAN is modular where each module mainly corresponds to a specific analysis
task. 

\citet{bommarito_empirical_2021} studied how packages contribute
dependencies to their dependents, but aggregated by developers. Very
interestingly, many of the most depended-on packages are maintained by
the same developers. They found the top 10 developers are responsible
for more than 50\% of all packages in the ecosystem, and the percent
would become higher if transitive dependency relations are also
considered. This suggests that the R ecosystem is more vulnerable on the
developer's level. 

\citet{German} separated R packages
into base packages, recommended packages, popular packages and
contributed packages where the order of the four groups reflects the
priority to be core packages in the ecosystem. They found that the more
core a package is, the more dependents it has. Additionally, they
explored other aspects such as code size, documentation, and community
interest. One interesting finding was code size for base packages keeps
increasing over years while in other categories code size is almost
stably unchanged. 

As there are also a huge amount of R packages hosted
on GitHub, \citet{Decan_inter} studied the inter-repository package
dependencies between CRAN and GitHub. They performed survival analysis
and found that packages on GitHub are easier to break due to updates of
their dependencies from CRAN. This may imply R packages on GitHub are
less maintained and tested. 

\citet{Mora_evolution} studied the
evolution of the R ecosystem. They revealed that in general packages are
stably updated over the years, but the number of packages as well as the
complexity of dependency relations are increasing. The increasing
complexity is expected as a result of the scale-free property of the
dependency graph where hub packages will preferably be linked to more
new packages when the ecosystem evolves.

There are also studies focusing on other aspects of the R ecosystem.
\citet{Claes} analyzed package errors on CRAN. They
found Solaris and MacOS have more errors mainly because they are less
used in development and packages are less tested on the two platforms.
They also found the majority of the errors on CRAN are from
external factors thus irrelevant to developers. 

\citet{Korkmaz2019}
measured the impacts of packages by the numbers of downloads, and they
studied how various factors can predict package impacts by utilizing a
generalized linear regression model. They used three groups of predictor
variables: 1. package features such as number of commits and GitHub
stars, 2. dependency networks centrality such as degree and closeness,
3. metrics from the co-developer
network\footnote{In a co-developer network, developers are nodes and two developers are connected if they contribute to the same package.}.
They found the number of dependents is a major factor that well
correlates to package impacts. Another interesting finding in this
study was if an author contributes to more packages, i.e., if an author is more active, 
his packages tend to have higher impacts.

In a very recent paper \citep{Vidoni}, the author called for more
software engineering studies on R programming language. She proposed
various futural directions for both software engineers and R developers
for better understanding the R ecosystem from the system's level.

\subsection{Current studies on other
ecosystems}\label{current-studies-on-other-ecosystems}

Given parent-child dependency relations of packages, the construction of
dependency graphs is similar for other programming languages. Methods
applied in one ecosystem can almost be seamlessly applied to other
ecosystems. We briefly described some of the current analyses and
findings as follows.

\citet{Decan_topo} compared topology of dependency graphs in
CRAN, npm (package repository of JavaScript) and PyPI (package
repository of Python) and they found various network metrics are
statistically different between ecosystems. This may reflect specific
patterns of how packages are implemented in different programming
languages. In particular, they found Python packages are more isolated
where a large number of them only depend on the standard core packages.

\citet{abate} studied the transitive dependency in the ecosystem.
They proposed a metric ``sensitivity'' which measures the total number
of downstream packages. They applied it to the Debian ecosystem and they
revealed packages with small amounts of direct dependents but affecting
majority of other packages in the ecosystem in an indirect manner.
They claimed sensitivity is a more meaningful metric for measuring risks
in the ecosystem. 

Evolution of a package involves bug fixing, adding new
features and changing interfaces, thus it may cause code breaks to
downstream packages. \citet{Bogart} studied different responses to
break changes, i.e., package updates that are not compatible with its
dependents packages, in the Eclipse, npm and CRAN ecosystems. They found
that different responses reflect the difference on policies and values
of the repositories, such as that Eclipse requires backward compatibility,
npm allows break changes via parallel versioning of the same package,
and CRAN requires compatibility always for the newest versions. 

\citet{Jafari} 
discussed strategies applied in dependent packages for
getting rid of break changes from upstream. By analyzing the JavaScript
ecosystem, they discovered seven bad habits for handling dependencies,
which they named as ``dependency smells'', for example, depending on a
fixed version, or depending directly on a remote repository with a URL.
They argued these smells will produce potential risks to the ecosystem
in the future. As a note, the dependency smell they categorized can
mostly be applied to other ecosystems.

For a package \emph{P}, \citet{abate} proposed a metric ``strong
dominance'' which measures the proportion of \emph{P}'s dependencies
that are from an upstream package \emph{Q}. The metric helps to identify
a package that dominantly contributes dependencies to its downstream
package \emph{P}. It is similar to the dependency heaviness metric we
applied in this paper, but the dependency heaviness metric is more
focused on the unique dependencies that \emph{Q} contributes to \emph{P}
in the ecosystem.

\section{Background concepts}\label{background}

Reusable code in R is formatted as packages. R provides a flexible way
for handling dependencies. In this section, we first introduce the
definitions of various dependency relations. Next we categorize packages
according to their relations in the dependency graph, and specific
heaviness metrics will be defined for them later in Section
\ref{heaviness_metrics}.

\subsection{Dependency relations}\label{dependency-relations}

\begin{figure}[htp]
{\centering \includegraphics[width=1\linewidth]{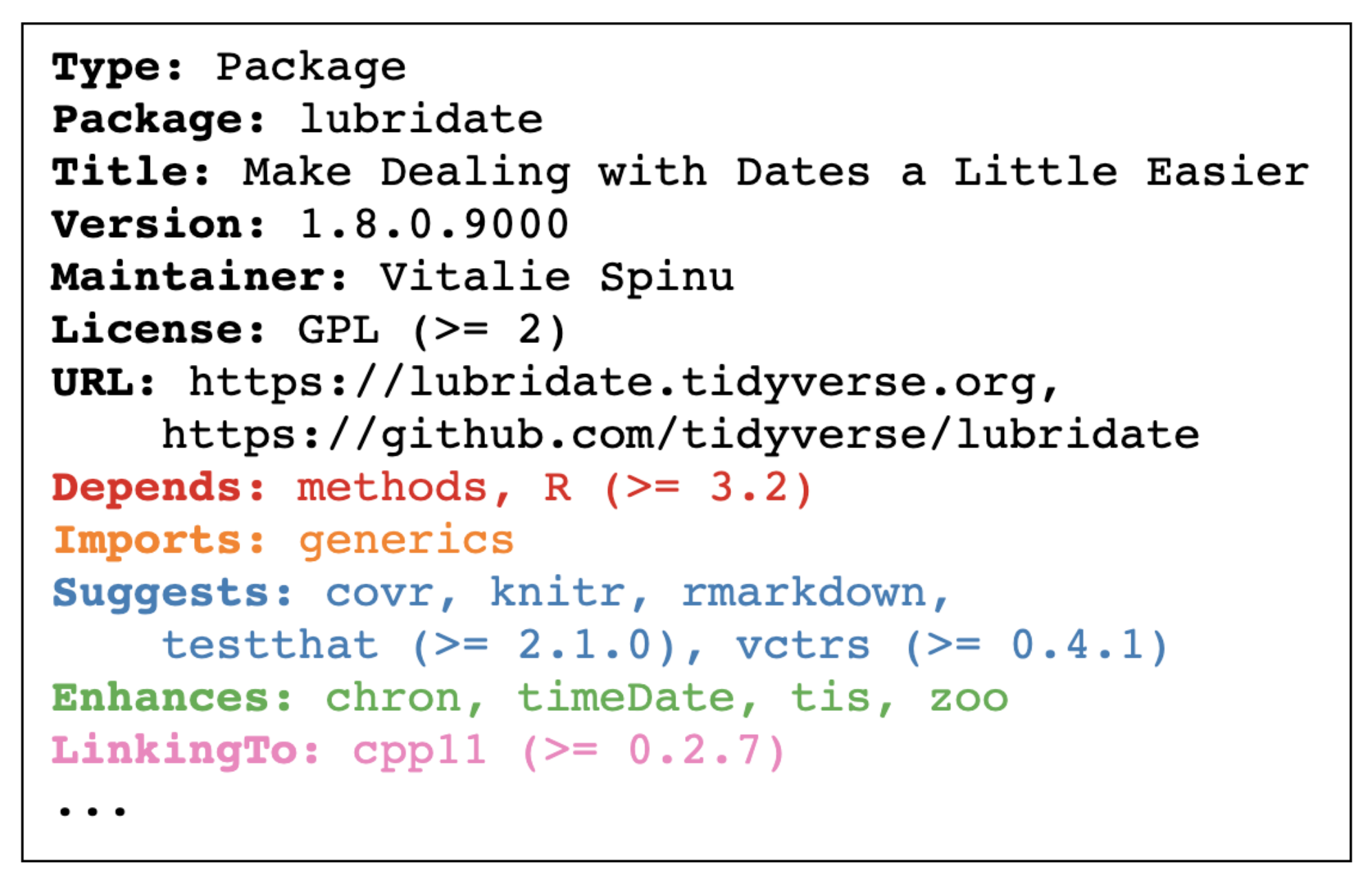}}
\vspace*{-5mm}
\caption{The DESCRIPTION file of the package \textit{lubridate}. Only a fragment of the file is demonstrated. This is an example where all five dependency fields are specified. For most of the packages in CRAN/Bioconductor, only a subset of them is specified. Version requirements can also be specified for dependency packages.}\label{fig:fig_01}
\end{figure}

\begin{figure*}
{\centering \includegraphics[width=1\linewidth]{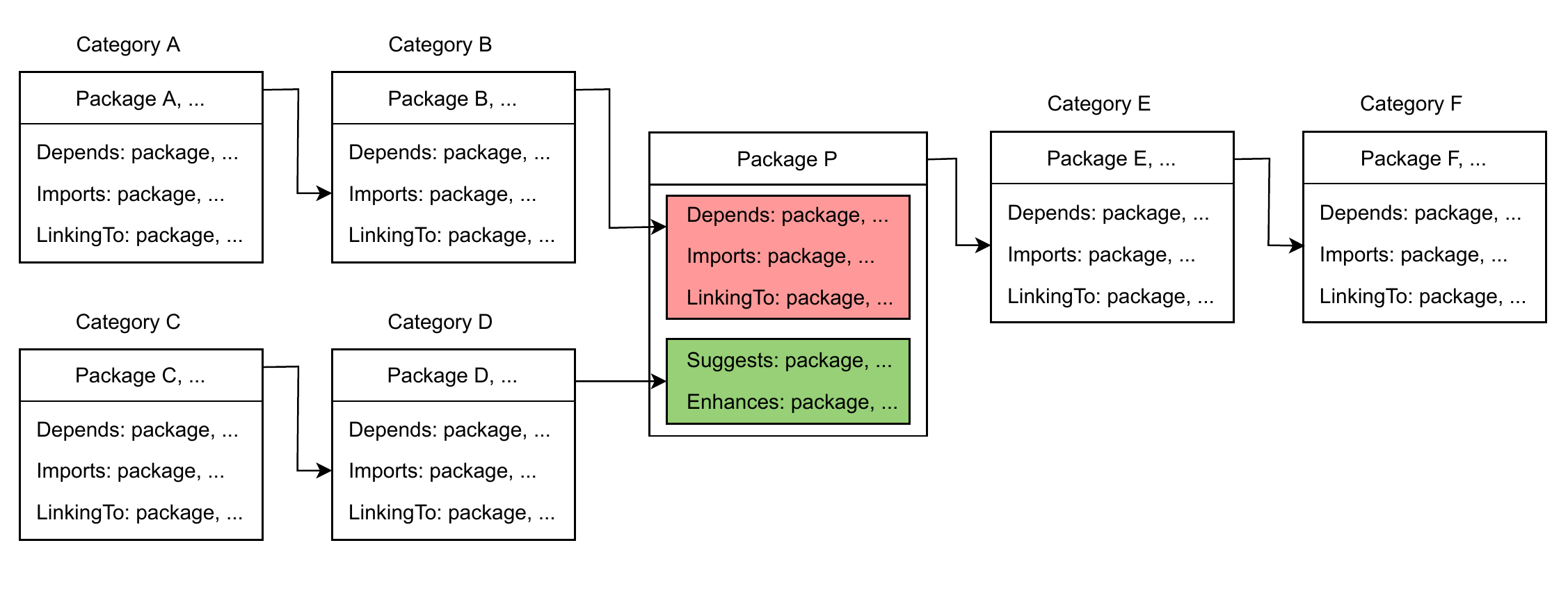}}
\vspace*{-10mm}
\caption{Demonstration of dependency relations between R packages. Packages in categories A-D are in the upstream of package \textit{P}. Packages in categories E and F are in the downstream of \textit{P}.}\label{fig:fig_02}
\end{figure*}

Dependency relations are declared in the fields of ``Depends'',
``Imports'', ``LinkingTo'', ``Suggests'' and ``Enhances'' in an R
package's DESCRIPTION file, which locates under the root directory of a
package (Figure \ref{fig:fig_01}). Denote a package as \emph{P},
dependency packages in its ``Depends'' are expected to be directly used
by users if \emph{P} is to be used, and they provide base
functionalities for \emph{P}. For example, it might be a good idea to
set the package \emph{ggplot2} as a ``Depends'' package for its
extension packages. Dependency packages in ``Depends'' are attached to
the search
path\footnote{Search path is an ordered list of environments and package namespaces where a function is sequentially looked up when it is executed by a user. The R function \texttt{search()} returns the search path.}
in the R session ahead of \emph{P} when executing the command \texttt{library(P)}.
All their public functions are visible to users. Dependency packages
listed in ``Imports'' are internally used by \emph{P} where specific
functions, methods or classes from there are imported into \emph{P}'s
namespace based on the rules defined in \emph{P}'s NAMESPACE file. The
``Imports'' packages are also loaded in the R session, but they are not
attached to the search path, thus not visible to users. Dependency
packages listed in ``LinkingTo'' contain header files to compile
\emph{P}'s C/C++ code. Packages listed in ``Depends'', ``Imports'' and
``LinkingTo'' are necessary for using \emph{P} and they must be
installed before installing \emph{P}. 

Fields ``Suggests'' and ``Enhances'' are similar. They contain dependency packages that are not
necessary for using \emph{P}, e.g., only used in examples or vignettes,
or in the code that provides optional functionalities of \emph{P}. Thus,
these two types of dependencies are not mandatory to be installed when
installing \emph{P}. However, dependencies in ``Suggests'' are by
default required for a complete
\texttt{R CMD check}\footnote{\texttt{R CMD check} is a command that performs comprehensive checks on a package. It checks source code, documentations, examples, vignettes and it runs unit tests. A successful check is required for acceptance on CRAN/Bioconductor.}.
``Enhances'' field is more flexible and packages listed are never
checked.

In the CRAN/Bioconductor ecosystem, the proportions of the five types of
dependency relations are 8.8\% for ``Depends'', 54.4\% for ``Imports'',
2.8\% for ``LinkingTo'', 33.7\% for ``Suggests'', and 0.4\% for
``Enhances'' from 194,351 relations in 22,076
packages\footnote{The dependency relations of all CRAN/Bioconductor packages were obtained with the R function \texttt{available.packages()}. Data was collected on 2022-06-08.}. Readers please refer to the official R manual
``Writing R Extensions'' \citep{R} for more details on the dependency relations. 

\subsection{Flexible control of dependencies}\label{flexible-control-of-dependencies}

Being different from other programming languages, R provides a flexible
control of dependencies. For example, in JavaScript, all dependencies
should be available in advance of using a library, even if the
functionality of a dependency is only rarely used by users. As a
comparison, R allows specifying a set of core dependencies as well as a
set of weak dependencies. Dependencies used in \emph{P}'s source code
can be explicitly specified in a form of \texttt{pkg::function(...)}
where \texttt{pkg} is a parent of \emph{P}. In this way,
\texttt{function()} is not directly imported from \texttt{pkg}, thus
availability of \texttt{pkg} is not mandatory for \texttt{R CMD check}
while it is only checked when \texttt{pkg::function(...)} is executed in
\emph{P}. Although \texttt{pkg} contributes to \emph{P}'s source code,
it is used optionally. Thus it can be declared as a weak dependency and
put in \emph{P}'s ``Suggests'' field. This actually provides the
possibility to optimize \emph{P}'s dependencies. When a parent of
\emph{P} brings a large number of extra dependencies but it only
provides limited functionalities which are rarely used by users, it is
reasonable to specify it as a weak dependency in
``Suggests''\footnote{Developer can write helper code to check whether weak parent is already installed. If not, a friendly message can be printed to inform users  to install it.}.
Actually this is the motivation of us to develop the \emph{pkgndep}
package which helps developers easily identify parent packages that
contribute heavy dependencies.

\begin{figure*}[!ht]
{\centering \includegraphics[width=0.95\linewidth]{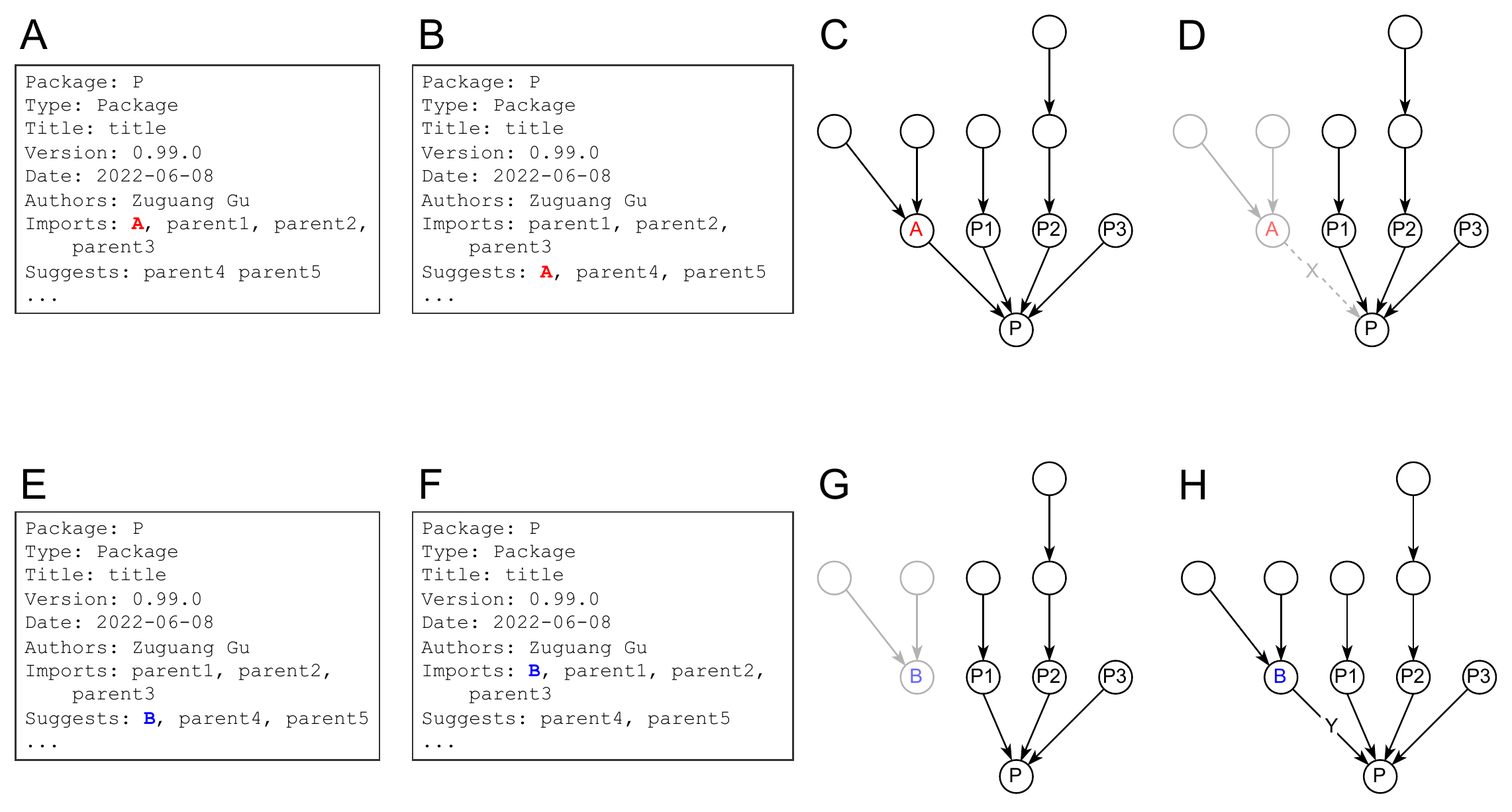}}
\caption{Demonstration of the heaviness definition. A) Fragment of the DESCRIPTION file of package \textit{P} where package \textit{A} is a strong parent. B) Change \textit{A} to a weak parent of \textit{P}, i.e., by moving \textit{A} to \textit{P}'s ``Suggests''. C) The dependency graph of all upstream packages of \textit{P}. Note the graph only contains strong dependency relations. D) Remove the relation between \textit{A} and \textit{P} from the dependency graph to simulate \textit{A} has become a weak parent of \textit{P}. E) Fragment of the DESCRIPTION file of package \textit{P} where package \textit{B} is a weak parent. F) Change \textit{B} to a strong parent of \textit{P}, i.e., by moving \textit{B} to \textit{P}'s ``Imports''. G) The dependency graph of all upstream packages of \textit{P}. Note since \textit{B} is a weak parent of \textit{P}, there is no connection between \textit{B} and \textit{P}. H) Add a new connection between \textit{B} and \textit{P} to simulate \textit{B} has become a strong parent of \textit{P}.}\label{fig:fig_03}
\end{figure*}

\subsection{Dependency categories}\label{dependency_category}

Depending on different dependency relations, the following dependency
categories for package \emph{P} are defined as follows.

\textbf{Strong parent packages}: Dependency packages listed in the
``Depends'', ``Imports'', and ``LinkingTo'' fields of \emph{P} (red box
in Figure \ref{fig:fig_02}). They are also called \textbf{strong direct
dependency packages} of \emph{P}. Strong parent packages are mandatory
to be installed when installing \emph{P}. To make it easy to discuss, we
always referred to them as \textbf{parent packages} in the paper.

\textbf{Weak parent packages}: Dependency packages listed in the
``Suggests'' and ``Enhances'' fields of \emph{P} (green box in Figure
\ref{fig:fig_02}). They are optionally required when installing
\emph{P}.

\textbf{Strong dependency packages}: Total dependency packages by
recursively looking for parent packages (category A, B, as well as
packages in red box in Figure \ref{fig:fig_02}). They are also called
\textbf{upstream packages}. Note strong dependency packages include
parent packages. Strong dependency packages are mandatory to be
installed when installing \emph{P} and failure of any strong dependency
package will prevent installation of \emph{P}. In some other studies,
they are also called \textbf{transitive dependency packages} \citep{decan_empirical_2019, Mora_CNA, Kikas}.

\textbf{All dependency packages}: Total dependencies by recursively
looking for parent packages, but on the level of \emph{P}, its weak
parents are also included (package category A, B, C and D, plus all
packages listed in the red and green boxes in Figure \ref{fig:fig_02}).
It simulates when the full functionality of \emph{P} is required, or
when all weak parents become strong parents, the total number of strong
dependency packages that \emph{P} requires. In this paper, we did not
discuss packages in this category, but for completeness, its definition
is still given here.

\textbf{Child packages}: Packages whose parents include \emph{P}
(category E in Figure \ref{fig:fig_02}). They are the packages on which
\emph{P} has a direct impact of dependencies. In some studies and
package repositories, they are also called \textbf{direct dependents} \citep{German}.

\textbf{Downstream packages}: Total packages by recursively looking for
child packages (category E and F in Figure \ref{fig:fig_02}). \emph{P}
is required for the installation of any of its downstream packages. Note
downstream packages include child packages.

\textbf{Indirect downstream packages}: Downstream packages excluding
child packages (category F in Figure \ref{fig:fig_02}), i.e., these with
distance to \emph{P} of at least 2 in the global dependency graph. These
are the packages on which \emph{P} has an indirect influence of
dependencies.

Except explicitly clarifying, the term ``dependency'' always refers to 
``strong dependency'' in the paper.

\section{Methods}\label{methods}

\subsection{Definitions of dependency
heaviness}\label{heaviness_metrics}

The dependency heaviness in general measures the number of additional
dependencies that a parent uniquely brings to its child packages or
downstream packages in the ecosystem. Depending on different dependency
categories defined in Section \ref{dependency_category}, there are
various heaviness metrics listed in the following subsections.

\textbf{Heaviness from a parent.} If package \emph{A} is a strong parent
of \emph{P}, the heaviness of \emph{A} on \emph{P} denoted as \(h\) is
calculated as

\begin{equation}
  \label{eq:1}
h = n_1 - n_2 
\end{equation}

\noindent where \(n_1\) is the number of strong dependencies of \emph{P}
(Figure \ref{fig:fig_03}A), and \(n_2\) is the number of strong
dependencies of \emph{P} after changing \emph{A} from a strong parent to
a weak parent, i.e., by moving \emph{A} to \emph{P}'s ``Suggests''
(Figure \ref{fig:fig_03}B). Thus, the heaviness measures the number of
additionally required strong dependencies that \emph{A} brings to
\emph{P} and are not brought by any other parent. In some discussions in
the paper, we explicitly denoted it as \(h^{A \rightarrow P}\) to
indicate the parent-child relation, or \(h^A\) if only the parent is of
interest.

If package \emph{B} is a weak parent of \emph{P}, \(n_2\) is defined as
the number of strong dependencies of \emph{P} after changing \emph{B} to
a strong parent of \emph{P}, i.e., by moving \emph{B} to \emph{P}'s
``Imports'' (Figure \ref{fig:fig_03}E-F). In this scenario, the
heaviness of the weak parent is calculated as \(n_2 - n_1\).

From the aspect of dependency graph denoted as a directed graph
\(G = (V, E)\) where \(V\) is the set of all packages and \(E\)
is the set of all strong parent-child dependency relations in the ecosystem, the heaviness of
\emph{A} on \emph{P} is a score associated with an edge
\(e^{A \rightarrow P} \in E\). Now \(n_1\) is the number of upstream
packages of \emph{P} (\(n_1\) = 9 in Figure \ref{fig:fig_03}C) and
\(n_2\) is the number of upstream package of \emph{P} after removing the
connection of \(A \rightarrow P\) from \(G\) (\(n_2\) = 6 in Figure
\ref{fig:fig_03}D), thus the heaviness of \emph{A} on \emph{P} is
\(n_1 - n_2 = 3\). When \emph{B} is a weak parent of \emph{P}, \(n_2\)
is calculated as the number of upstream packages after adding a new
connection of \(B \rightarrow P\) to \(G\) (Figure \ref{fig:fig_03}G-H).

Since weak parents are not necessarily required for \emph{P}, in this
paper, we only discussed the heaviness from strong parents.

\textbf{Max heaviness from parents.} Normally, a package has multiple
parents. Max heaviness from parents helps to reveal the parent that
dominantly brings extra dependencies to package \emph{P}. Assume
\emph{P} has \(K_{\mathrm{p}}\) parents, the heaviness denoted as
\(h_{\mathrm{max}}\) is defined as

\begin{equation}
  \label{eq:2}
h_{\mathrm{max}}=\underset{k\in\{1..K_{\mathrm{p}}\}}{\max}h_{k}
\end{equation}

\noindent where \(h_k\) is the heaviness of the \textit{k}\textsuperscript{th} parent on \emph{P}.

\textbf{Heaviness from an upstream package.} If package \emph{C} is an
upstream package of \emph{P} in the global dependency graph, let \(n_1\)
be the number of strong dependencies of \emph{P}, and let \(n_2\) be the
number of strong dependencies of \emph{P} after changing \emph{C} to a weak
parent of all \emph{C}'s child packages, then the heaviness of \emph{C}
on \emph{P} denoted as \(h_{\mathrm{u}}\) is calculated as

\begin{equation}
  \label{eq:3}
h_{\mathrm{u}} = n_1 - n_2 .
\end{equation}

From the aspect of the dependency graph, \(n_2\) is the number of
upstream packages of \emph{P} after removing all edges which start from
\emph{C}. When \emph{C} is only a parent of \emph{P} (i.e., distance
from \emph{C} to \emph{P} is one), \(h_{\mathrm{u}}\) is not always
identical to \(h\). Assume \emph{C} is a parent of \emph{P} and \emph{P}
has another parent \emph{A} where \emph{C} is also a parent of \emph{A},
i.e., with the relations of \(C \rightarrow P\) and
\(C \rightarrow A \rightarrow P\). \(h\) just measures the local
dependency effect which only removes \(C \rightarrow P\), while
\(h_\mathrm{u}\) is a global dependency effect which removes all links
from \emph{C}, i.e., both \(C \rightarrow P\) and \(C \rightarrow A\),
which results in general \(h_\mathrm{u} \ge h\). In some discussions in
the paper, we denoted it explicitly as
\(h_\mathrm{u}^{C \rightarrow P}\) to indicate the upstream-downstream
relation.

\textbf{Heaviness on child packages.} Assume \emph{P} has
\(K_\mathrm{c}\) child packages and the \textit{k}\textsuperscript{th}
child is denoted as \(A_k\). Denote the number of strong dependencies of
\(A_k\) as \(n_{1k}\), and denote the number of strong dependencies of
\(A_k\) after changing \emph{P} to a weak parent of \(A_k\) as
\(n_{2k}\), the heaviness of \emph{P} on its child packages denoted as
\(h_\mathrm{c}\) is calculated as

\begin{equation}
  \label{eq:4}
h_\mathrm{c}=\frac{1}{K_\mathrm{c}}\sum_{k=1}^{K_\mathrm{c}}(n_{1k}-n_{2k}) .
\end{equation}

\(n_{1k} - n_{2k}\) is actually the heaviness of \emph{P} on \(A_k\), Equation \ref{eq:4}
can be rewritten as

\begin{equation}
  \label{eq:4-2}
h_\mathrm{c}=\frac{1}{K_\mathrm{c}}\sum_{k=1}^{K_\mathrm{c}}h^{P \rightarrow A_k} .
\end{equation}

The heaviness measures the average number of additional dependencies
that \emph{P} brings to its child packages. 

\textbf{Heaviness on downstream packages.} The definition is similar to
the heaviness on child packages. Assume \emph{P} has \(K_\mathrm{d}\)
downstream packages and the \textit{k}\textsuperscript{th} downstream
package is denoted as \(B_k\). Denote the number of strong dependencies
of \(B_k\) as \(n_{1k}\), and denote the number of strong dependencies of
\(B_k\) after changing \emph{P} to a weak parent of all \emph{P}'s child
packages as \(n_{2k}\). The heaviness of \emph{P} on its downstream
packages denoted as \(h_\mathrm{d}\) is calculated as

\begin{equation}
  \label{eq:5}
h_\mathrm{d}=\frac{1}{K_\mathrm{d}}\sum_{k=1}^{K_\mathrm{d}}(n_{1k}-n_{2k}) .
\end{equation}

From the aspect of the dependency graph, \(n_{2k}\) is the number of
upstream packages of \(B_k\) in a reduced graph where \emph{P} is
removed. Equation \ref{eq:5} can be rewritten as

\begin{equation}
  \label{eq:5-2}
h_\mathrm{d}=\frac{1}{K_\mathrm{d}}\sum_{k=1}^{K_\mathrm{d}}h_\mathrm{u}^{P \rightarrow B_k}
\end{equation}

\noindent where \(h_\mathrm{u}^{P \rightarrow B_k}\) is the heaviness of \emph{P} as an upstream package on \(B_k\).
In this way, \(h_\mathrm{c}\) and \(h_\mathrm{d}\)
are not always identical if \emph{P} only has child packages. \(h_\mathrm{c}\) is a local measure while
\(h_\mathrm{d}\) is a global measure. They have the relation of
\(h_\mathrm{d} \ge h_\mathrm{c}\).

\textbf{Heaviness on indirect downstream packages.} The calculation is
the same as \(h_\mathrm{d}\) except here child packages are excluded
from downstream packages. Denote the heaviness as \(h_\mathrm{id}\) and
denote the set of \emph{P}'s child packages as \(S_\mathrm{c}\),
\(h_\mathrm{id}\) is defined as

\begin{equation}
  \label{eq:6}
h_\mathrm{id}=\frac{1}{K_\mathrm{d}-K_\mathrm{c}}\sum_{k=1}^{K_\mathrm{d}}(n_{1k}-n_{2k})\cdot I(B_k\notin S_\mathrm{c})
\end{equation}

\noindent where \(K_\mathrm{c}\) and \(K_\mathrm{d}\) are the numbers of
child and downstream packages respectively, and \(I()\) is an indicator
function. \(h_\mathrm{id}\) is set to 0 if
\(K_\mathrm{c} = K_\mathrm{d}\), i.e., \emph{P} has no indirect
downstream packages. \(h_\mathrm{id}\) measures the contribution of
dependencies of \emph{P} to the ecosystem in an indirect way.

\begin{figure*}[htp]
{\centering \includegraphics[width=0.95\linewidth]{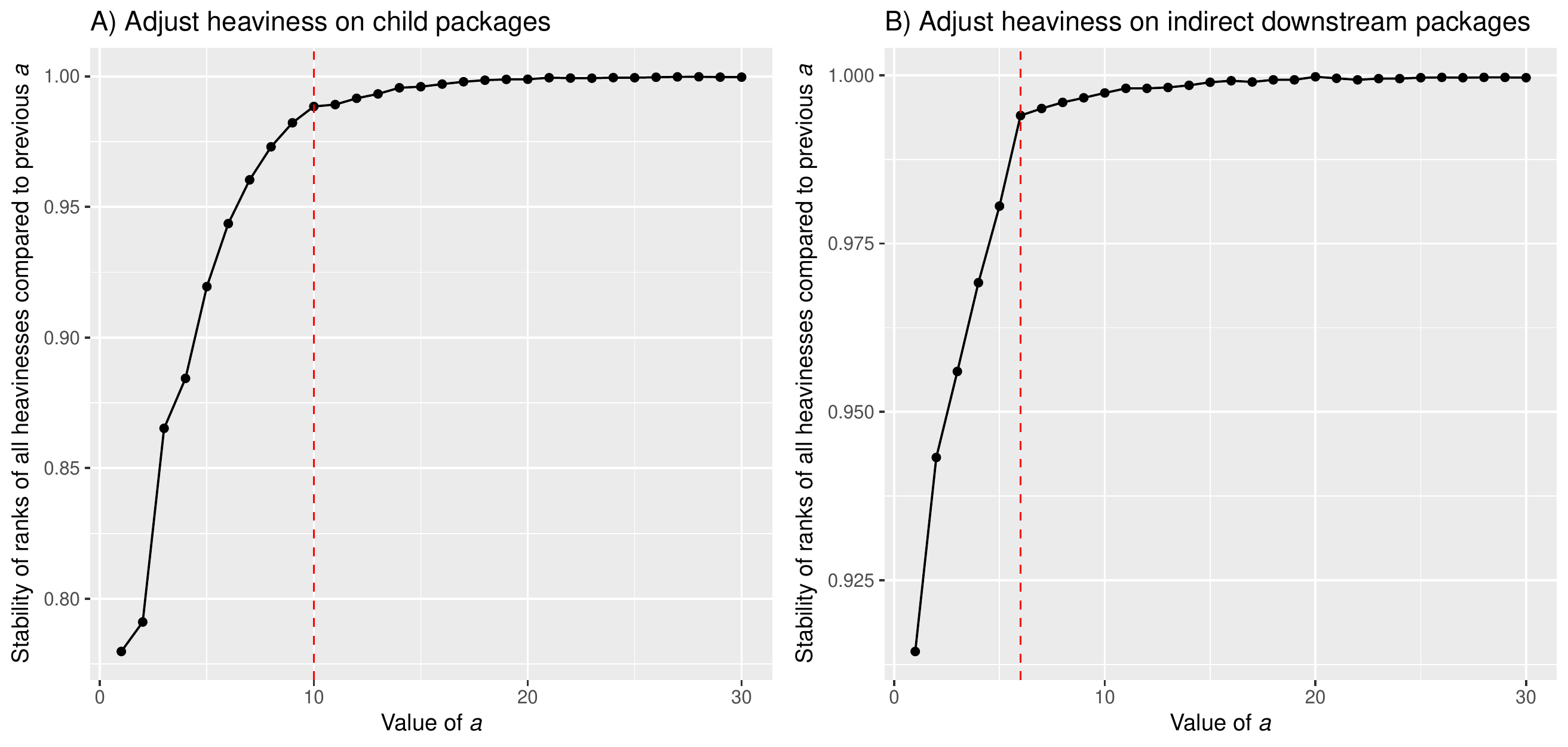}}
\caption{Select a proper penalty value \textit{a} for adjusting heaviness values. A) Adjust heaviness on child packages. B) Adjust heaviness on indirect downstream packages. Vertical dashed lines are the final selections of the penalty values.}\label{fig:fig_04}
\end{figure*}

\subsection{Adjusted heaviness}\label{adjusted-heaviness}

In the Results section (Section \ref{results}), we performed dependency
heaviness analysis on the CRAN/Bioconductor ecosystem. One of the aims
is to prioritize packages which are significantly affected by upstream
packages or affect their downstream packages in the ecosystem. If
grouping packages by \(K\) which can be the number of parent, child or
downstream packages depending on different types of heaviness metrics,
distributions of heaviness values always have long tails, and tails are
especially longer for smaller \(K\) (Figure \ref{fig:fig_06} and
\ref{fig:fig_09}). Thus, if simply ranking packages based on the
original heaviness values, top packages are preferably associated with
small \(K\). In general, packages with small \(K\) are of less interest
because they only have very small impacts on the ecosystem. To
prioritize packages with broader impacts on the ecosystem, the original
definitions of various heaviness metrics are adjusted to decrease the
weights of packages with smaller \(K\). Please note, the designs of the
adjusted heaviness metrics are empirical and the absolute values of
adjusted heaviness are meaningless, which are only used for ranking
packages.

\textbf{Adjusted max heaviness from parents.} When a package has more
parents, dependencies from individual parents would have more overlap
(i.e., dependencies from parent \emph{A} overlap to dependencies
from parent \emph{B}). Since heaviness only measures the number of
unique dependencies that a single parent brings in, or in other words,
the number of dependencies that are mutually exclusive to those brought
by all other parents, with more parents, the max heaviness from parents
would decrease. For a package indexed as \(k\) in the ecosystem, the
original max heaviness from parent denoted as \(h_{\mathrm{max},k}\) is
adjusted to \(h_{\mathrm{max},k}^{\mathrm{adj}}\), by multiplying a
zooming factor denoted as \(a_k\):

\begin{equation}
  \label{eq:7}
h_{\mathrm{max},k}^{\mathrm{adj}}=a_k\cdot h_{\mathrm{max},k} .
\end{equation}

\(a_k\) is defined as

\begin{equation}
  \label{eq:8}
a_k=(n_k+30)/n_{\mathrm{max}}
\end{equation}

\noindent where \(n_k\) is the number of parents of the
\textit{k}\textsuperscript{th} package, and \(n_{\mathrm{max}}\) is the
maximal number of parents of all packages in the ecosystem. The value of
30 was selected empirically to balance the zooming rate on different
\(n_k\).

\textbf{Adjusted heaviness on child packages.} Generally, heaviness on
child packages has a trend that distribution tails are shortened when
the numbers of child packages increase (Figure \ref{fig:fig_09}). This
is mainly because if package \emph{P} has more child packages, its child
packages may have more other parents which dilute the heaviness from
\emph{P}. To decrease the weights of packages with small numbers of
child packages, a positive penalty term denoted as \(a\) is added to
\(K_\mathrm{c}\) as in Equation \ref{eq:9} where \(K_{\mathrm{c}}\) is
the number of child packages and \(h_{\mathrm{c}}^{\mathrm{adj}}\) is
the adjusted heaviness of a package on its child packages. Note \(a\)
is set to the same value for all packages.

\begin{equation}
  \label{eq:9}
h_\mathrm{c}^{\mathrm{adj}}=\frac{1}{K_\mathrm{c}+a}\sum_k^{K_\mathrm{c}}(n_{1k}-n_{2k})=\frac{K_\mathrm{c}}{K_\mathrm{c}+a}\cdot h_\mathrm{c} .
\end{equation}

It is easy to see that \(a\) decreases \(h_\mathrm{c}\) faster for
smaller \(K_\mathrm{c}\) than larger \(K_\mathrm{c}\). To select an
optimized value for \(a\), we took \(a\) as integers in the set
\(\{1, 2, \dotsc, 29, 30\}\); and for a specific package indexed as
\(k\) and a value of \(a\), we calculated the adjusted heaviness on its
child packages denoted as \(h_{\mathrm{c},k,a}^{\mathrm{adj}}\), and the
vector for all packages is denoted as
\(h_{\mathrm{c},a}^{\mathrm{adj}}\). \(a\) is selected as the value by
which the ranking of adjusted heaviness of all packages becomes stable.
To measure the stability of the ranking in
\(h_{\mathrm{c},a}^{\mathrm{adj}}\) compared to
\(h_{\mathrm{c},a-1}^{\mathrm{adj}}\), we calculated the stability score
denoted as \(s_a\) as

\begin{figure*}[!ht]
{\centering \includegraphics[width=0.95\linewidth]{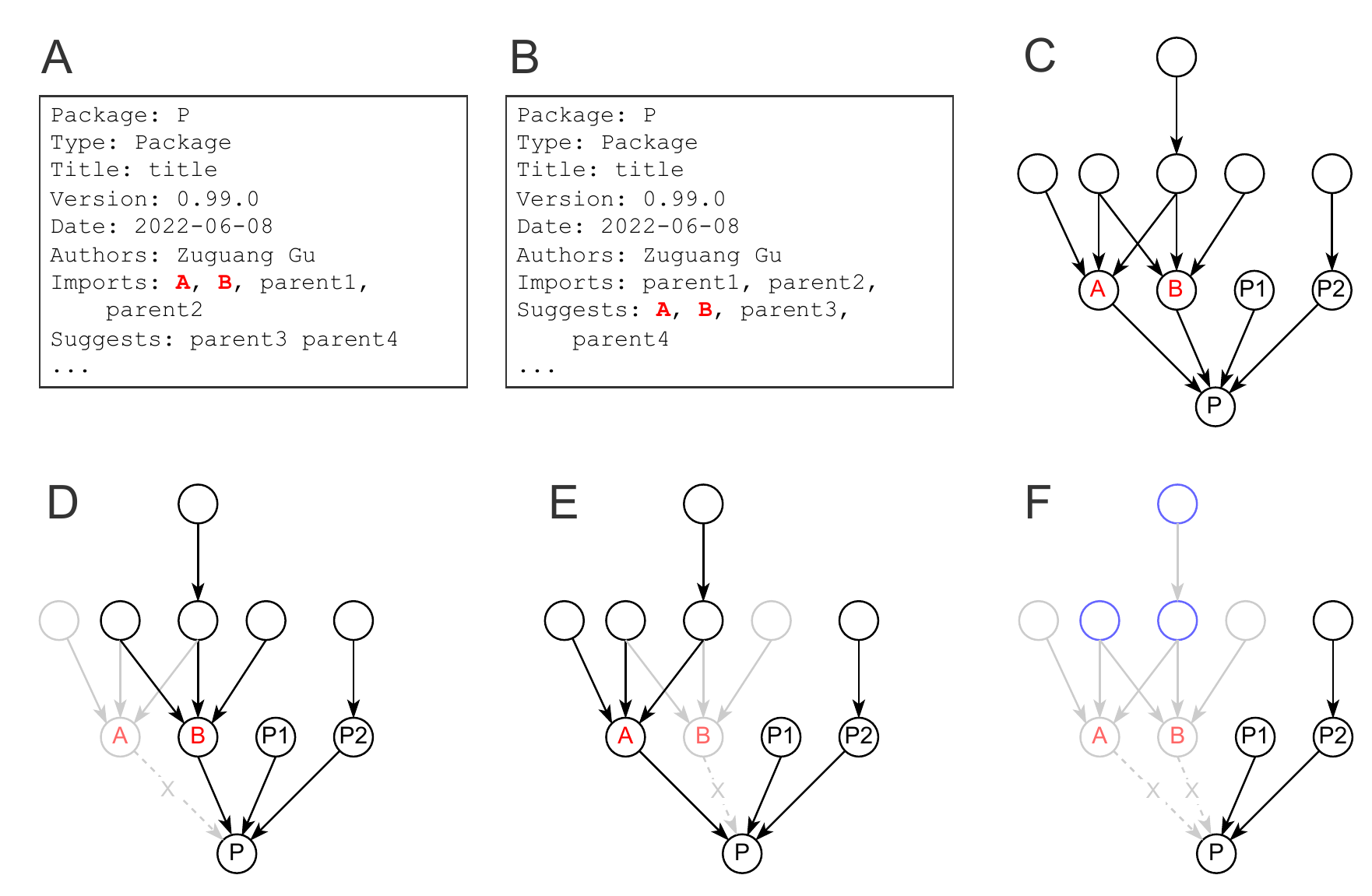}}
\caption{Demonstration of the co-heaviness definition. A) Fragment of the DESCRIPTION file of package \textit{P} where package \textit{A} and \textit{B} are two strong parents. B) Change both \textit{A} and \textit{B} to weak parents of \textit{P}, i.e., by moving \textit{A} and \textit{B} to \textit{P}'s ``Suggests''. C) The dependency graph of all upstream packages of \textit{P}. Note the graph only contains strong dependency relations. D) Remove the relation between \textit{A} and \textit{P} from the dependency graph to simulate \textit{A} has become a weak parent of \textit{P}. E) Remove the relation between \textit{B} and \textit{P} from the dependency graph to simulate \textit{B} has become a weak parent of \textit{P}. F) Remove the relations from both \textit{A} and \textit{B} to \textit{P} from the dependency graph to simulate \textit{A} and \textit{B} have become two weak parents of \textit{P}. The number of blue nodes in Figure F corresponds to the value of co-heaviness of \textit{A} and \textit{B} on \textit{P}.}\label{fig:fig_05}
\end{figure*}

\begin{equation}
  \label{eq:10}
s_a=\frac{1}{N}\sum_k^NI(|R_{k,a}-R_{k,a-1}|\le50)
\end{equation}

\noindent where \(N\) is the total number of packages in the ecosystem,
\(R_{k,a}\) and \(R_{k,a-1}\) are the ranks of package \(k\)'s adjusted
heaviness in the two vectors \(h_{\mathrm{c},a}^{\mathrm{adj}}\) and
\(h_{\mathrm{c},a-1}^{\mathrm{adj}}\) respectively, and \(I()\) is the
indicator function.

\(s_a\), or its general denotation \(s\), measures the fraction of
packages whose ranking differences of adjusted heaviness are no larger
than 50 between two neighboring values of \(a\) (50 is a small value
compared to the total number of R packages in the ecosystem, which is
22,076 in this study). When \(s\) becomes stable with \(a\), we could
conclude increasing \(a\) won't greatly change the ranking in \(s\). In
Figure \ref{fig:fig_04}A, we calculated \(s\) on a list of \(a\) in
\(\{1, 2,\dotsc, 29, 30\}\). By observing the trend of the curve, we can
see when \(a = 10\) (red vertical in Figure \ref{fig:fig_04}A), \(s\)
starts to increase slowly. Thus, \(a\) was empirically selected to 10.

\textbf{Adjusted heaviness on indirect downstream packages.} The
definition is the same as the adjusted heaviness on child packages. For
a package \emph{P}, the adjusted heaviness denoted as
\(h_\mathrm{id}^\mathrm{adj}\) is calculated as:

\begin{equation}
  \label{eq:11}
h_\mathrm{id}^\mathrm{adj}=\frac{K_\mathrm{id}}{K_\mathrm{id}+a}\cdot h_\mathrm{id}
\end{equation}

\noindent where \(K_\mathrm{id}\) is the number of indirect downstream packages,
i.e., \(K_\mathrm{id} = K_\mathrm{d} - K_\mathrm{c}\). The penalty \(a\)
was selected in a similar way as for child packages. It was selected as
\(a = 6\) (Figure \ref{fig:fig_04}B).

The heaviness on all downstream packages can be adjusted in the same
way. However, in this study, we only looked at the heaviness on indirect
downstream packages instead of total downstream packages, thus, here we
omitted the definition of adjustment of heaviness on all downstream
packages.

\subsection{Co-heaviness from parent pairs}\label{coheaviness}

Heaviness from parent \emph{A} on \emph{P} only measures the number of
additional dependencies that \emph{A} uniquely brings to \emph{P}.
However, there are scenarios where multiple parents import a similar set
of dependencies, which results in heaviness from individual parents
being very small. Here we define the co-heaviness that measures the
number of additional dependency packages simultaneously brought by two
parent packages (Figure \ref{fig:fig_05}). Let \emph{A} and \emph{B} be
two parents of \emph{P}, let \(S_A\) be the set of reduced dependency
packages when only changing \emph{A} to a weak parent of \emph{P}
(Figure \ref{fig:fig_05}D), let \(S_B\) be the set of reduced
dependency packages when only changing \emph{B} to a weak parent of
\emph{P} (Figure \ref{fig:fig_05}E), and let \(S_{AB}\) be the set of
reduced dependency packages when changing both \emph{A} and \emph{B} to
weak parents of \emph{P}, then the co-heaviness of \emph{A} and \emph{B}
on \emph{P} denoted as \(h_\mathrm{co}\) is defined as

\begin{equation}
  \label{eq:12}
h_\mathrm{co}=\left|S_{AB}\setminus\cup(S_A,S_B)\right|
\end{equation}

\noindent where \(X \setminus Y\) is the set of elements in \(X\) but
not in \(Y\), and \(|X|\) is the number of elements in set \(X\) (Figure
\ref{fig:fig_05}F). The co-heaviness measures the number of reduced
packages only caused by co-action of \emph{A} and \emph{B}. In some
discussions in the paper, we denoted it explicitly as
\(h_\mathrm{co}^{(A,B)\rightarrow P}\) to indicate the relations.

With the previous denotations, there are

\begin{equation}
  \label{eq:13}
h^A = |S_A|
\end{equation} \vspace*{-7mm}

\begin{equation}
\label{eq:14}
h^B = |S_B|
\end{equation} \vspace*{-7mm}

\begin{equation}
\label{eq:15}
S_A \cap S_B = \emptyset
\end{equation}

\noindent where \(h^A\) is the heaviness of \emph{A} on \emph{P},
\(h^B\) is the heaviness of \emph{B} on \emph{P}, and \(S_A\) and
\(S_B\) are mutually exclusive. Then there is the following relation:

\begin{equation}
  \label{eq:16}
|S_{AB}| = h_\mathrm{co} + h_A + h_B .
\end{equation}

Thus, the number of reduced dependencies by moving both \emph{A} and
\emph{B} to \emph{P}'s weak parents is the sum of heaviness of \emph{A}
and \emph{B} on \emph{P} individually and the co-heaviness of \emph{A}
and \emph{B} on \emph{P} (Figure \ref{fig:fig_05}F).

\section{Tools and materials}\label{data}

In our previous study \citep{pkgndep}, we have developed an R
package \emph{pkgndep} which performs dependency heaviness analysis on
single packages. For a given R package whose dependency packages are
only from CRAN and Bioconductor, \emph{pkgndep} calculates the
dependency heaviness from every of its parent package, additionally with
an intuitive heatmap visualization as well as an HTML report, which
helps developers to easily reveal heavy parents. To be a companion tool
for the study described in this paper, we have updated \emph{pkgndep} to
version 1.2.0 with new functionalities for analyzing the R package
ecosystem\footnote{\url{https://CRAN.R-project.org/package=pkgndep}.}. Given a
package \emph{P}, there are the following functions for querying package
dependencies in various dependency categories:

\begin{itemize}
\item
  \texttt{parent\_dependency()}
\item
  \texttt{upstream\_dependency()}
\item
  \texttt{child\_dependency()}
\item
  \texttt{downstream\_dependency()}
\end{itemize}

And there are the following functions for calculating the corresponding
dependency heaviness metrics:

\begin{itemize}
\item
  \texttt{heaviness()}
\item
  \texttt{co\_heaviness()}
\item
  \texttt{heaviness\_from\_upstream()}
\item
  \texttt{heaviness\_on\_children()}
\item
  \texttt{heaviness\_on\_downstream()}
\end{itemize}

The function names are self-explanatory. We believe these new functions
will be convenient tools for researchers who wish to perform software
engineering studies on the R package ecosystem.

The analysis on the R package ecosystem is bound to a certain snapshot
of CRAN and Bioconductor. In this study, dependency relations of all
CRAN and Bioconductor packages (we call it the ``package database'')
were obtained with the function \texttt{available.packages()} on
2022-06-08\footnote{\texttt{available.packages()} always returns the metadata of the newest versions of all packages hosted on CRAN/Bioconductor.}.
This resulted in 18,638 R packages from CRAN, 3,438 packages from
Bioconductor (bioc version 3.15), and 124,251 strong dependency relations in the two ecosystems. Various dependency analyses in this
study were applied with the \emph{pkgndep} package with its
aforementioned new functions. The result is represented as a table where
rows are R packages and columns are various heaviness metrics. The table
can be obtained by the function \texttt{all\_pkg\_stat\_snapshot()} in
\emph{pkgndep}. Network analysis was applied with the package
\emph{igraph} \citep{igraph} and visualized with Cytoscape
\citep{cytoscape} and the R package \emph{RCy3} \citep{RCy3}.

For reproducibility of this study, we have integrated the script for
calculating dependency metrics for all packages in the \emph{pkgndep}
package and it can be accessed with the command
\texttt{system.file("extdata",\ "analysis.R",\ package\ =\ "pkgndep")}.
The scripts for the figures in this paper are available at
\url{https://github.com/jokergoo/pkgndep_global}. \emph{pkgndep} version 1.2.* can be used to reproduce the complete analysis in this study. We plan to regularly update
the package database in \emph{pkgndep} to ensure the dependency analysis
on R packages is always up-to-date.

\begin{table*}
\caption{\label{tab:table1}Average values of various metrics of packages on CRAN and Bioconductor. $N_\mathrm{child}$: number of child packages; $N_\mathrm{indirect}$: number of indirect downstream packages.}
\centering
\begin{tabular}[t]{lrr}
\toprule
Metrics averaged in the ecosystem & CRAN & Bioconductor \\
\midrule
Number of strong dependencies & 30.8 & 66.1 \\
Number of parents & 5.1 & 8.4 \\
Max heaviness from parents & 13.3 & 24.6 \\
Max co-heaviness from parents & 4.5 & 12.2 \\
Number of children & 4.7 & 3.5 \\
Number of children (with $N_\mathrm{child} > 0$) & 18.2 & 15.2 \\
Heaviness on child packages (with $N_\mathrm{child} > 0$) & 7.8 & 14.8 \\
Number of indirect downstream & 29.0 & 11.5 \\
Number of indirect downstream (with $N_\mathrm{indirect} > 0$) & 256.8 & 136.5 \\
Heaviness on indirect downstream packages (with $N_\mathrm{indirect} > 0$) & 4.4 & 8.3 \\
\bottomrule
\end{tabular}
\end{table*}

\section{Research questions}\label{rq}

Dependency heaviness is a directional measure. It measures the amount of
dependencies uniquely transmitted from a package to its single or total
downstream packages. In the context of the complex dependency graph, we
separated our research questions (RQs) into three categories according to
different dependency directions.

Packages have certain numbers of parents and each parent may bring
additional dependencies transitively. We studied the dependency flow
from parent to child packages and we first asked the following two
research questions:

\begin{itemize}
\item
  \textbf{RQ1: What is the general pattern of dependency heaviness from
  parents?} This includes the following sub questions: 1. What is the
  proportion of packages suffering from heavy parents and what are they?
  2. How is the dependency heaviness accumulated from remote upstream
  packages?
\item
  \textbf{RQ2:} It is very common that a package depends on multiple
  parents. The second research question is \textbf{how do two parents
  contribute dependencies synergistically?}
\end{itemize}

For hub packages such as \emph{ggplot2} in the ecosystem, their
dependencies are all passed to their downstream packages. With
regards to the impacts on downstream packages, we asked the next
following two research questions:

\begin{itemize}
\item
  \textbf{RQ3: What is the general pattern of dependency heaviness of a
  package contributing to all its child packages?} This also includes
  several sub questions: 1. What is the distribution of dependency
  heaviness on child packages? 2. What are the top packages that spread
  the highest amount of dependencies directly to their children?
\item
  \textbf{RQ4:} A package may transmit dependencies to its remote
  downstream packages. We next asked \textbf{what are the differences between
  the dependency heaviness transmission to indirect downstream packages
  and to direct child packages?}
\end{itemize}

CNA on the dependency graph can reveal interesting structures of
dependency transmission in the ecosystem. From the aspect of network
analysis, we asked the last research question:

\begin{itemize}
\item
  \textbf{RQ5. How are the dependency flows transmitted throughout the
  ecosystem?} This includes the following sub questions: 1. What is the
  global attribute and structure of the dependency graph if taking
  dependency heaviness as weight? 2. How deep can the dependency
  heaviness be transmitted? 3. Does there exist a core graph that
  transmits the majority of the dependencies in the ecosystem?
\end{itemize}

When answering these research questions, we also examined the difference between
 CRAN and Bioconductor to reveal ecosystem-specific patterns.

\section{Results}\label{results}

\subsection{RQ1: Heaviness from parent
packages}\label{heaviness_from_parents}

A package may have multiple parents. Here we only studied the max
heaviness from its parents (abbreviated as MHP) which measures the
number of unique dependencies that a package maximally inherits from its
parents. In Section \ref{uniqueness_heavy_parent}, we demonstrated
that if a package suffers heavy dependency from its parents, it is very
likely that one or only a very few parents contribute heavy dependencies
to it. Therefore, MHP is a reasonable metric for studying the general
pattern of heaviness from parent packages.

\begin{figure*}[htp]
{\centering \includegraphics[width=0.95\linewidth]{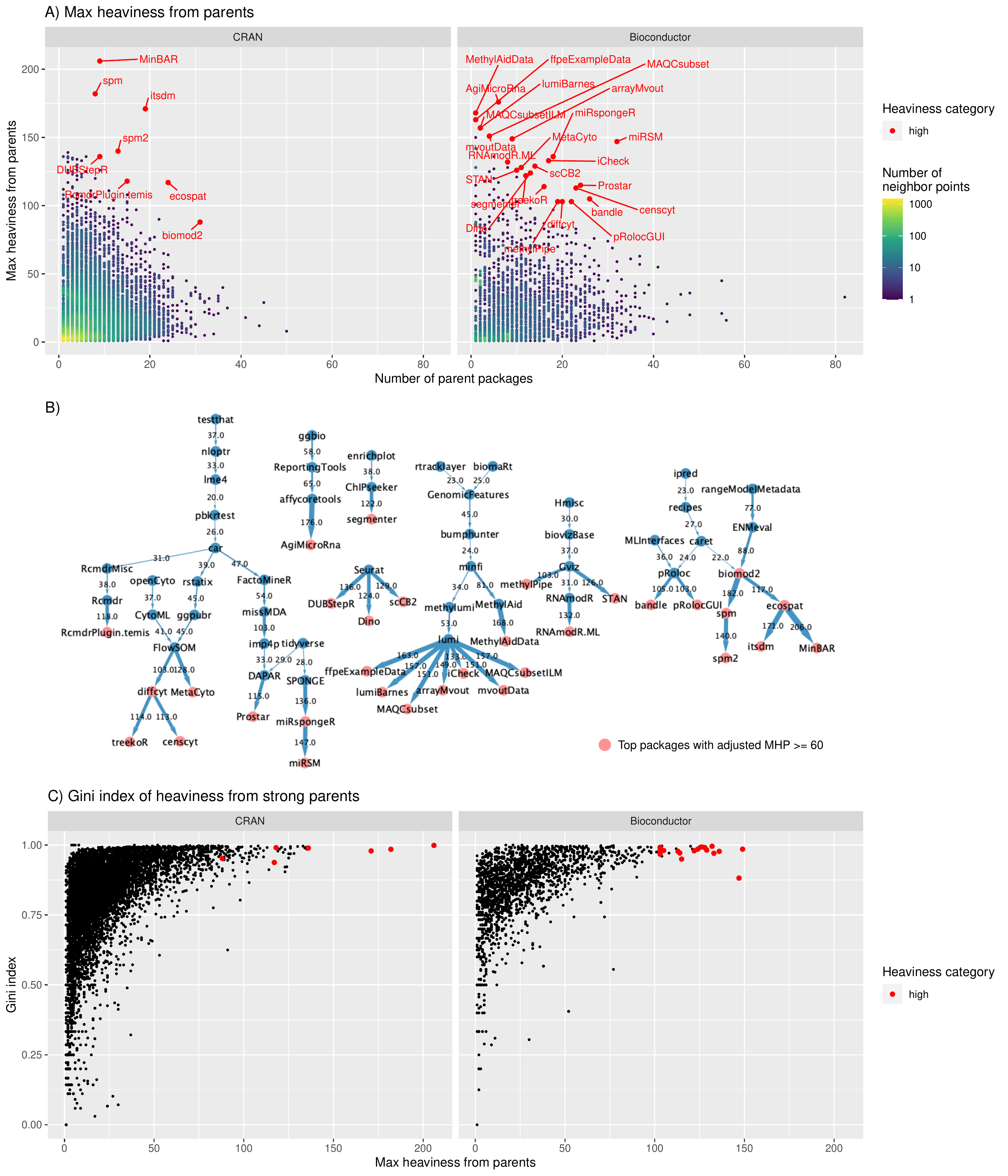}}
\caption{Heaviness from parent packages. A) Distribution of MHP. B) Upstream of packages with adjusted MHP $\ge$ 60. Only edges with heaviness $\ge$ 20 are included. Values on the edges are the dependency heaviness of a parent on its child package. C) Gini index of the heaviness from parents. In Figure A, since points may overlap, we visualized the two-dimensional distribution densities where numbers of neighbor points were based on circular areas with radius of 1\% of ranges on x-axis and y-axis respectively. The packages highlighted in red in Figure A are the same as in Figure B and C, which are the top packages with adjusted MHP $\ge$ 60. MHP: max heaviness from parents.}\label{fig:fig_06}
\end{figure*}

\subsubsection{Distribution of MHP}\label{distribution-of-mhp}

In general, MHP has a long-tail distribution where 82.4\% of all
packages have a heaviest parent contributing heaviness \textless{} 30,
whereas there are only 878 (4.0\%) packages having a heaviest parent
contributing heaviness \(\ge\) 60. Therefore, only a small fraction of
packages have heavy parents in the ecosystem. For 878 packages with
extremely heavy parents (MHP \(\ge\) 60), 785 (89.4\%) of them only have
less than 15 parents, which indicates packages with small numbers of
parents are more likely to have heavy parents. Also interestingly, 720
(82.0\%) of them do not have child packages, which indicates when
packages have heavier parents, they are more likely to be leaf packages
in the dependency graph and no other package depends on them.

Distribution tails of MHP are shortened when numbers of parents increase
(Figure \ref{fig:fig_06}A). As has been explained in Section
\ref{heaviness_metrics}, MHP measures the number of unique dependencies
that a single parent maximally brings in. Thus, with more parents,
dependencies from multiple parents would have more chances to overlap,
which results in decreasing MHP. In Figure \ref{fig:fig_06}A, it can be
straightforwardly observed that, on the top edge of the point clouds,
there is a clear trend where MHP drops as the number of parents
increases.

Globally, Bioconductor packages have heavier parents than CRAN packages.
The mean MHP for all Bioconductor packages is 24.6 while it is only 13.3
for all CRAN packages (Table \ref{tab:table1}. The median values are 16
vs 6). Nevertheless, the difference of MHP in the two repositories
becomes smaller when the number of parents increases. E.g., when only
considering packages with parents \textgreater{} 20 (580 packages left),
the mean MHP values are 27.2 vs 22.8 for CRAN and Bioconductor packages,
and the median values are 23 vs 19.

Bioconductor packages suffer heavier parents than CRAN packages. The
reason might be that Bioconductor packages are mainly for biological
data analysis and many of them integrate various analyses and annotation
resources from upstream packages (average numbers of parents are 5.1 and
8.4 for CRAN and Bioconductor, Table \ref{tab:table1}). Thus, it is
easier for dependency heaviness to be accumulated from upstream on Bioconductor
(average numbers of strong dependencies are 30.8 and 66.1 for CRAN and
Bioconductor, Table \ref{tab:table1}). For example, the Bioconductor package
\emph{miRspongeR} listed as a top package in Figure \ref{fig:fig_06}A
has an upstream package \emph{SPONGE} which integrates analysis from a
list of heavy parents for different analysis aims (\emph{tidyverse} for
data processing, \emph{ggpubr} for visualization, \emph{biomaRt} for
obtaining biological annotation data, \emph{caret} for statistical
modeling).

\subsubsection{Top packages with the highest
MHP}\label{top-packages-with-the-highest-mhp}

If a package has more upstream dependencies, it is more vulnerable to
corruptions from its upstream packages. We identified top packages with
the highest MHP and these top packages have the most risky parents in
the ecosystem from the aspect of how they uniquely contribute risks to
their child packages.

Packages with smaller numbers of parents have longer distribution tails
of MHP. To capture top packages with the heaviest parents but not biased
by the small numbers of parents, the original MHP was adjusted. Top
packages that have extremely heavy parents are filtered by adjusted MHP
\(\ge\) 60 and they are marked in red in Figure \ref{fig:fig_06}A. We
found there are 8 CRAN packages and 24 Bioconductor packages that have
extremely heavy parents. This also indicates Bioconductor packages may
suffer more from heavy parents. Among them, the package \emph{MinBAR}
has the highest MHP with the heaviest parent named \emph{ecospat} which
uniquely contributes 206 additional dependencies to \emph{MinBAR}.

When a package suffers from an extremely heavy parent, the next question
naturally to ask is how are the dependencies accumulated from the
upstream of the heavy parent? Figure \ref{fig:fig_06}B illustrates
upstream dependencies of the top 32 packages with adjusted MHP \(\ge\)
60. To keep the graph small and compact, parent-child with dependency
heaviness \(\ge\) 20 are only included in the graph. In other words, the
graph in Figure \ref{fig:fig_06}B contains major dependency flows from
upstream to the packages suffering heavy parents. Interestingly, for
most packages, high dependencies are accumulated from upstream in
very long ranges. The longest transmission path in the graph has a
length of 9, e.g., from \emph{testthat} to \emph{treekoR} or \emph{censcyt}. As a
comparison, the average distance in the global dependency graph is only
2.6. We also found heavy dependencies can be inherited from the same
parents, such as \emph{Seurat} transmitting on average 130 unique
dependencies to its 3 child packages and \emph{lumi} transmitting on
average 152 unique dependencies to its 7 child packages. The
transmission of heaviness will be further discussed in Section
\ref{heaviness_on_children} on the short-range transmissions and in
Section \ref{cna_analysis} on the long-range transmissions.

\subsubsection{Uniqueness of the heaviest
parents}\label{uniqueness_heavy_parent}

A package may have multiple parents. We observed a general pattern
that only a small number of parents contribute large heaviness while the
majority of other parents only contribute very small heaviness to their
child packages. The Gini index was used to quantitatively measure the
dispersion of the heaviness distribution from a package's parents. When
the Gini index is close to 1, there is a uniquely high heaviness value;
and when the Gini index is close to zero, the heaviness values
approximately follow a uniform distribution. Figure \ref{fig:fig_06}C
shows that there is a clear trend that when MHP increases, i.e., when
the parents get heavier, the Gini indices increase as well. For the top
packages with extremely high MHP, the corresponding Gini indices are
very close to 1. This indicates these heaviest parents play unique and
dominant roles in contributing dependencies to their child packages.

\vspace*{3mm}
\begin{tcolorbox}[colbacktitle=black!20!white,colback=white,coltitle=black,boxrule=0.5pt,sharpish corners,title=\textit{Answer to RQ1},left=1mm,right=1mm,top=1mm,bottom=1mm]
Only a small fraction of R packages in the ecosystem suffer from heavy
parents. The heavy dependencies on them are normally accumulated from
far upstream. If a package suffers from heavy parents, it is very likely
the heaviest parent plays a unique and dominant role in contributing
dependencies to it.
\end{tcolorbox}

\subsection{RQ2: Co-heaviness from two parent
packages}\label{rq2-co-heaviness-from-two-parent-packages}

Heaviness discussed in Section \ref{heaviness_from_parents} only
measures the number of dependencies that a single parent uniquely brings
in. However, there are scenarios where multiple parents import similar
sets of dependencies, which results in heaviness from individual parents
being very small. Taking the package \emph{DESeq2} (version 1.36.0) as an example, its
two parent packages \emph{geneplotter} and \emph{genefilter} import 51
and 53 dependencies respectively, among which 50 are the
same\footnote{The heaviness analysis on \textit{DESeq2} can be accessed in the heaviness database introduced in Section \ref{database}.}.
Due to the high overlap, the heaviness of \emph{geneplotter} and
\emph{genefilter} on \emph{DESeq2} are only 1 and 2. In this section, we
studied the dependency heaviness simultaneously imported by two strong
parents, i.e., the co-heaviness. In the analysis, for a package, we only
looked at the maximal co-heaviness from all its parent pairs
(abbreviated as MCoHP). Empirically, MCoHP is dramatically higher than
the co-heaviness from other pairs if a package has an obvious heavy
parent pair. For the ease of discussion, we name the parent pair which
contributes the highest co-heaviness as the \emph{MCoHP parents} in the
paper.

\subsubsection{Relations of parent
pairs}\label{relations-of-parent-pairs}

\begin{figure}[t]
{\centering \includegraphics[width=1\linewidth]{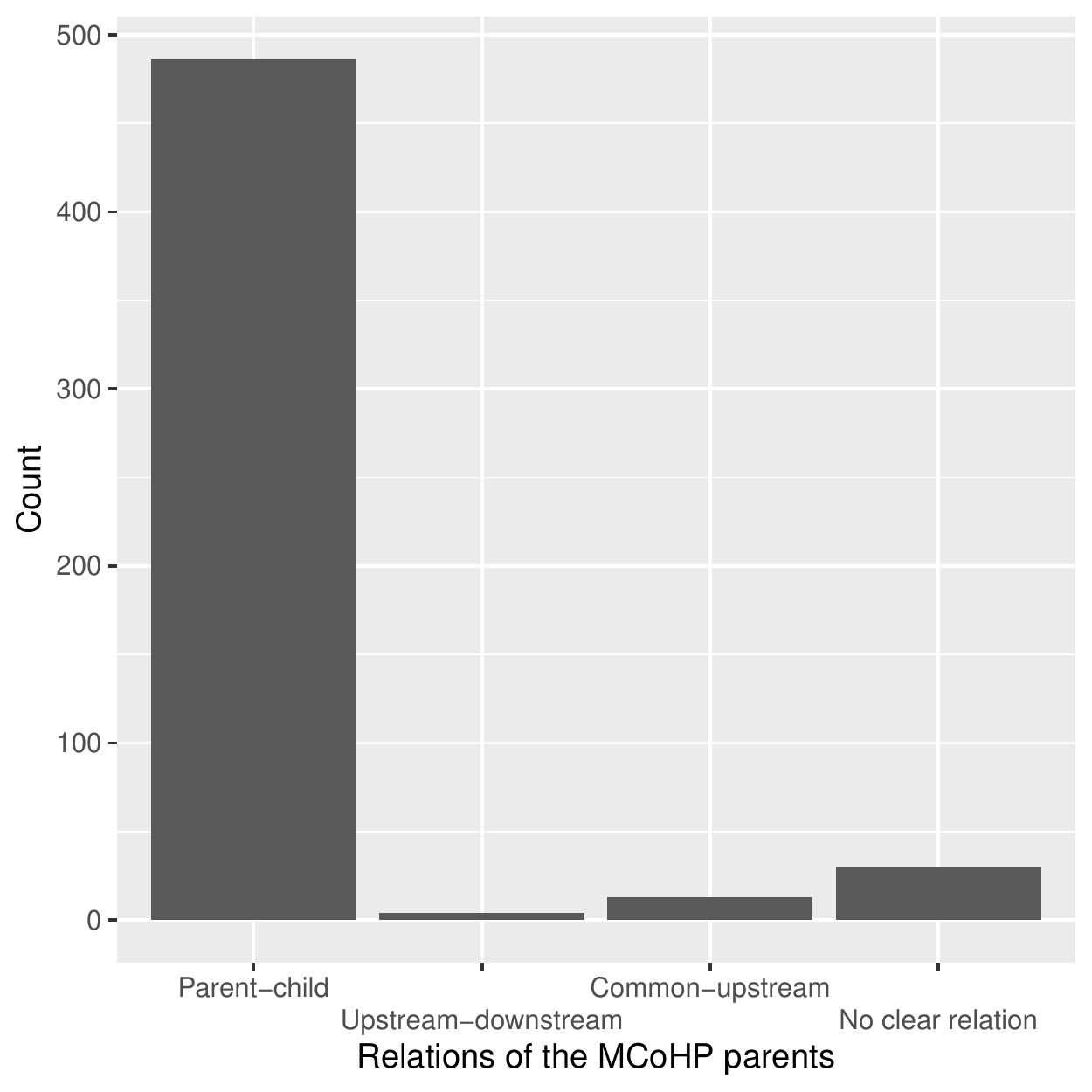}}
\vspace*{-5mm}
\caption{Relations of the MCoHP parents of the 533 packages with MCoHP $\ge$ 40. MCoHP: max co-heaviness from parents.}\label{fig:fig_07}
\end{figure}

In the CRAN/Bioconductor ecosystem, we found 92.6\% of all packages have
MCoHP \textless{} 20, and only 533 (2.4\%) of them (143 from CRAN and
390 from Bioconductor) have MCoHP \(\ge\) 40, which indicates the
proportion of packages whose heavy dependencies are simultaneously
inherited from two parents are extremely small in the ecosystem. Denote two
parents of package \emph{P} as \emph{A} and \emph{B}, the relations of
parent pairs can be summarized into the following four categories:

\begin{enumerate}
\def\labelenumi{\arabic{enumi}.}
\item
  \textbf{Parent-child}. E.g., \emph{B} is a parent of \emph{A} where
  \emph{A} inherits all dependencies from \emph{B}.
\item
  \textbf{Upstream-downstream}. E.g., \emph{B} is an indirect upstream
  package of \emph{A} where \emph{A} also inherits all dependencies from
  \emph{B}.
\item
  \textbf{Common-upstream}. There exists a common upstream package
  \emph{C} of \emph{A} and \emph{B} where \emph{C} contributes heavy
  dependencies to both \emph{A} and \emph{B}, defined as
  \(h_\mathrm{u}^{C\rightarrow A}>0.75h_\mathrm{co}^{(A,B)\rightarrow P}\)
  and
  \(h_\mathrm{u}^{C\rightarrow B}>0.75h_\mathrm{co}^{(A,B)\rightarrow P}\).
  This means the co-heaviness of \emph{A} and \emph{B} mainly comes from
  \emph{C}.
\item
  \textbf{No clear relation}. The dependencies of \emph{A} and \emph{B}
  are accumulated from their upstream packages independently; or
  \emph{A} and \emph{B} may have common upstream, but the upstream
  packages do not contribute significantly high heaviness on them.
\end{enumerate}

For the 533 top packages with MCoHP \(\ge\) 40, in 486 (91.2\%) packages,
MCoHP parents are in parent-child relation; in 4 (0.75\%) packages,
MCoHP parents are in upstream-downstream relation; in 13 (2.4\%)
packages, MCoHP parents are in common-upstream relation; and in 30
(5.6\%) packages, MCoHP parents have no clear relation (Figure
\ref{fig:fig_07}).

\begin{figure*}[htp]
{\centering \includegraphics[width=0.95\linewidth]{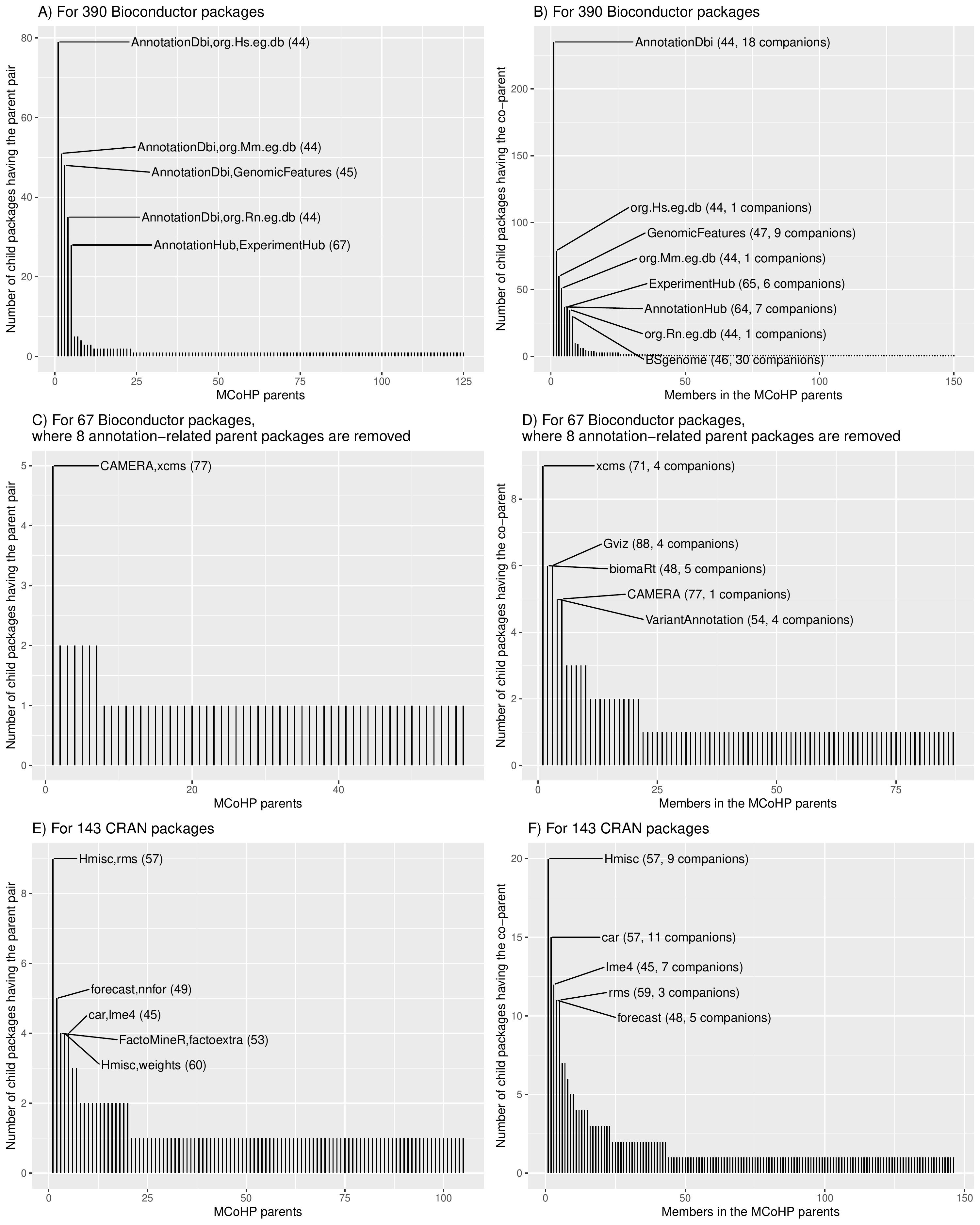}}
\caption{Co-heaviness in the CRAN/Bioconductor ecosystem. A, C, E) Number of child packages for each MCoHP parent pair. A: all 390 top Bioconductor packages; C: the top Bioconductor packages of which MCoHP parents do not include the annotation packages listed in Figure B; E: all 143 top CRAN packages. In Figure A, C, E, numbers in the parentheses are the average co-heaviness. B, D, F) Number of child packages for packages as a member of MCoHP parents. In Figure B, D, F, numbers in the parentheses are the average co-heaviness and numbers of companion packages. All analyses are restricted to 533 packages with MCoHP $\ge$ 40. MCoHP: max co-heaviness from parents.}\label{fig:fig_08}
\end{figure*}

\subsubsection{Parent pairs that contribute high
co-heaviness}\label{co_heaviness_pair}

Since most MCoHP parents are in parent-child relations, we next studied
whether there is preference of selecting a member of the
MCoHP parents. Again only in the 533 packages with MCoHP \(\ge\) 40, we
found in the Bioconductor ecosystem, the package \emph{AnnotationDbi}
dominantly contributes high co-heaviness to its child packages as a
MCoHP parent (Figure \ref{fig:fig_08}A-B). Top four companion packages
co-working with \emph{AnnotationDbi} are \emph{org.Hs.eg.db} (affecting
79 child packages), \emph{org.Mm.eg.db} (affecting 51 child packages),
\emph{GenomicFeatures} (affecting 48 child packages) and
\emph{org.Rn.eg.db} (affecting 35 child packages) (Figure
\ref{fig:fig_08}A). \emph{AnnotationDbi} is a base Bioconductor package
which defines the general database interface, and a large number of
other packages storing specific biological annotation data are
implemented with this interface. To use a specific annotation package
from Bioconductor, methods from \emph{AnnotationDbi} should also be used
for extracting corresponding data. Therefore, the data packages (e.g.,
\emph{org.Hs.eg.db} for human, \emph{org.Mm.eg.db} for mouse and
\emph{org.Rn.eg.db} for rat) and \emph{AnnotationDbi} are normally used
together. Similarly, \emph{AnnotationHub} and \emph{ExperimentHub}
provide similar functionalities for storing external annotation and
experimental data. \emph{ExperimentHub} inherits the same set of
function interface from \emph{AnnotationHub}, thus any package depending
on \emph{ExperimentHub} would also depend on \emph{AnnotationHub}. These
two ``\emph{*Hub}'' packages are becoming standard ways for handling
external data and they are used more and more frequently in the
Bioconductor system. Figure \ref{fig:fig_08}B shows the number of child
packages of the MCoHP parent packages. Besides the aforementioned three
``\emph{org.*.eg.db}'' packages, \emph{AnnotationDbi} co-works with in
total 18 packages on Bioconductor and it affects 235 child
packages with mean co-heaviness of 44. There are other examples of heavy
MCoHP parents (Figure \ref{fig:fig_08}B). The package
\emph{GenomicFeatures} co-works with 9 other packages where
\emph{GenomicFeatures} provides tools for manipulating gene/transcript
annotations and the companion packages are organism-specific annotation
packages. The package \emph{BSgenome} co-works with 30 other packages
where \emph{BSgenome} provides classes and methods for dealing
with genome sequences, and the companion packages are also
organism-specific annotation packages.

As has been demonstrated previously, top packages contributing strong
co-heaviness in the Bioconductor ecosystem are all annotation-related
packages. We next explored Bioconductor by excluding the eight top
annotation-related packages listed in Figure \ref{fig:fig_08}B. We
aimed to study the co-heavines patterns for the ``software packages'' on
Bioconductor. Figure \ref{fig:fig_08}C shows now there is almost no
dominant parent pair contributing strong co-heaviness, except the pair
\emph{CAMERA} and \emph{xcms}. These two packages are both for mass
spectrometry data analysis and they contribute a mean co-heaviness of 77
to their 5 child packages. \emph{CAMERA} and \emph{xcms} provide core
functionalities for mass spectrometry data analysis and additional
packages in the field might need to depend on both of them. Figure
\ref{fig:fig_08}D lists top Bioconductor software packages as heavy
co-parents. Notably, the package \emph{Gviz} which provides a general
visualization framework for genomics data affects 6 child packages as a
co-parent with other 4 comparison packages. It contributes on average a
co-heaviness of 88 which is much higher than other co-parent packages.

\begin{table*}

\caption{\label{tab:table2}Top packages with adjusted HC $\ge$ 30. $N_{\mathrm{strong}}$: number of strong dependencies; $N_{\mathrm{child}}$: number of child packages; HC: heaviness on child packages.}
\centering
\begin{tabular}[t]{lrrrl@{\hskip 0.5in}lrrrl}
\toprule
Package & $N_{\mathrm{strong}}$ & $N_{\mathrm{child}}$ & HC & Repository & Package & $N_{\mathrm{strong}}$ & $N_{\mathrm{child}}$ & HC & Repository\\
\midrule
ecospat & 232 & 3 & 151.0 & CRAN & GenomicScores & 98 & 26 & 56.0 & Bioconductor\\
RTCGA & 127 & 9 & 128.0 & Bioconductor & AER & 92 & 22 & 52.6 & CRAN\\
lumi & 162 & 13 & 114.2 & Bioconductor & WGCNA & 108 & 33 & 52.3 & CRAN\\
Rcmdr & 135 & 45 & 101.2 & CRAN & drc & 96 & 17 & 51.1 & CRAN\\
Deducer & 107 & 5 & 94.6 & CRAN & tidyverse & 107 & 89 & 48.4 & CRAN\\
\addlinespace
Seurat & 145 & 38 & 85.3 & CRAN & devtools & 76 & 80 & 47.0 & CRAN\\
taxize & 127 & 12 & 77.4 & CRAN & Gviz & 142 & 37 & 43.5 & Bioconductor\\
TraMineR & 100 & 7 & 77.1 & CRAN & FactoMineR & 104 & 52 & 41.2 & CRAN\\
smacof & 122 & 8 & 75.2 & CRAN & caret & 81 & 180 & 41.0 & CRAN\\
brms & 123 & 13 & 65.1 & CRAN & car & 87 & 183 & 40.6 & CRAN\\
\addlinespace
MESS & 84 & 9 & 63.8 & CRAN & ggpubr & 96 & 125 & 37.0 & CRAN\\
minfi & 141 & 38 & 62.4 & Bioconductor & rms & 78 & 54 & 36.8 & CRAN\\
survminer & 115 & 27 & 58.2 & CRAN & fda & 60 & 78 & 33.9 & CRAN\\
\bottomrule
\end{tabular}
\end{table*}

For CRAN packages, there are less package pairs that contribute strong
co-heaviness to their child packages (average MCoHP are 4.5 and 12.2 for
CRAN and Bioconductor, Table \ref{tab:table1}). There is only one parent
pair \emph{Hmisc} and \emph{rms} that contribute strong co-heaviness on
their 9 child packages with average co-heaviness of 57 (Figure
\ref{fig:fig_08}E). Also, \emph{Hmisc} affects 20 child packages as a
co-parent with other 9 companion packages (Figure \ref{fig:fig_08}F)
where \emph{Hmisc} is a parent of all its companion packages.

\vspace*{3mm}
\begin{tcolorbox}[colbacktitle=black!20!white,colback=white,coltitle=black,boxrule=0.5pt,sharpish corners,title=\textit{Answer to RQ2},left=1mm,right=1mm,top=1mm,bottom=1mm]
Only a small fraction of R packages in the ecosystem inherit
dependencies uniquely and simultaneously from two parents. The two
parents contributing high co-heaviness mostly have the relation of
parent-child, thus they provide a similar set of dependencies. There are
more Bioconductor packages having high co-heaviness from parents where
one parent provides an interface for manipulating biological data and
the other parent provides data for specific organisms with that
interface. On CRAN, there are very few packages suffering from high
co-heaviness from parents.
\end{tcolorbox}

\begin{figure*}[!htp]
{\centering \includegraphics[width=0.9\linewidth]{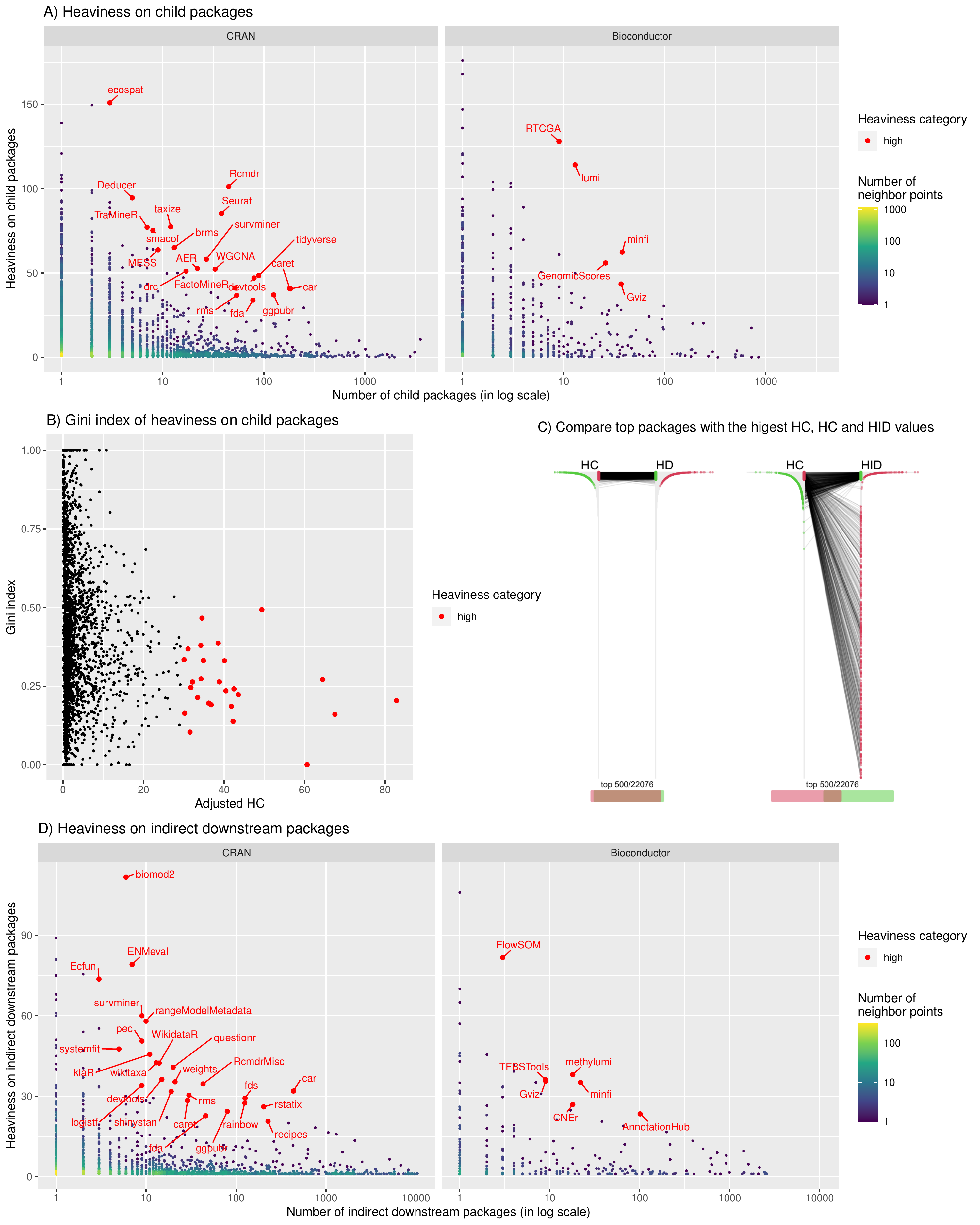}}
\caption{Heaviness analysis on child and indirect downstream packages. A) Heaviness on child packages. B) Gini indices of individual heaviness values on child packages. C) Compare top 500 packages with the highest HC, HD and HID values. In each plot, packages are sorted by two metrics, e.g., HC and HD in the left plot. Top 500 packages by either of the two metrics are highlighted by segments that connect the same packages in the two sorted lists. At the bottom is a Venn diagram showing the overlap of the top 500 packages from the two metrics. D) Heaviness on indirect downstream packages. HC: heaviness on child packages; HD: heaviness on downstream packages; HID: heaviness on indirect downstream packages.}\label{fig:fig_09}
\end{figure*}

\subsection{RQ3: Heaviness on child
packages}\label{heaviness_on_children}

We studied how a package \emph{P} contributes dependency heaviness to
its child packages (abbreviated as HC) by looking at the average number
of dependencies uniquely brought by \emph{P}.

\subsubsection{Distribution of HC}\label{distribution-of-hc}

Figure \ref{fig:fig_09}A illustrates the distribution of HC for 5,593
(25.3\%) packages with at least one child. In general, distributions of
HC have very long tails for packages with small numbers of children, and
distribution tails are shortened dramatically as numbers of children
increase. This is mainly because when \emph{P} has more child packages,
the heaviness on child packages may be diluted by their other parents.

On average, CRAN packages have slightly more children than Bioconductor
(18.2 vs 15.2, Table \ref{tab:table1}), but with smaller average HC (7.8
vs 14.8, Table \ref{tab:table1})\footnote{Statistics are based on packages with at least one child package.}. In Section
\ref{heaviness_from_parents}, we have demonstrated Bioconductor packages
have more parents and inherit more strong dependencies, thus it is
expected that they on average transmit more dependency heaviness to
their children.

\subsubsection{Top packages with the highest
HC}\label{top-packages-with-the-highest-hc}

From the perspective of R package developers, HC is especially useful
because it also measures the expected number of additional dependencies
if \emph{P} is added as a new strong parent of their packages. Thus it
is important to list top packages that contribute high heaviness in the
ecosystem. Similarly, we also adjusted original HC and we set adjusted
HC \(\ge\) 30 to extract top packages that broadly affect a large number
of child packages in the ecosystem (Figure \ref{fig:fig_09}A and Table \ref{tab:table2}). Not
surprisingly, these top packages already inherit large amount of
dependencies from their parents (the column \(N_\mathrm{strong}\) in Table
\ref{tab:table2}), and most of these dependencies are transmitted to
their child packages uniquely (the column \(N_\mathrm{child}\) in Table
\ref{tab:table2}). To list a few, the package \emph{lumi} has 162
dependencies from its parents, and on average 114 (70.4\%) of them are
uniquely contributed to its child packages; the package \emph{RTCGA} has
127 dependencies from its parents, and all of them uniquely go to its
child packages. Also as listed in Table \ref{tab:table2}, there are
several packages transmitting large heaviness to broader sets of child
packages, such as the package \emph{caret} contributing on average 41
additional dependencies to its 180 child packages, the package
\emph{car} contributing on average 40.6 additional dependencies to its
183 child packages, and the package \emph{ggpubr} contributing on
average 37.0 additional dependencies to its 125 child packages. These
packages play major roles in contributing heavy dependencies to their
child packages in the ecosystem.

\begin{table*}[t]
\caption{\label{tab:table3}Top packages with adjusted HC $\ge$ 30 or adjusted HID $\ge$ 20. Additionally packages listed in the table have numbers of downstream packages (i.e., numbers of child packages + numbers of indirect downstream packages) $\ge$ 30. $N_{\mathrm{child}}$: number of child packages; HC: heaviness on child packages; $N_{\mathrm{id}}$: number of indirect downstream packages; HID: heaviness on indirect downstream packages.}
\centering
\begin{tabular}{lrrrrcc}
\toprule
Package & $N_{\mathrm{child}}$ & HC & $N_{\mathrm{id}}$ & HID & Is it a top HC package? & Is it a top HID package?\\
\midrule
car & 183 & 40.6 & 435 & 31.9 & y & y\\
recipes & 26 & 9.9 & 227 & 20.6 &  & y\\
rstatix & 10 & 12.1 & 203 & 26.0 &  & y\\
caret & 180 & 41.0 & 29 & 28.4 & y & y\\
ggpubr & 125 & 37.0 & 80 & 24.4 & y & y\\
\addlinespace
AnnotationHub & 97 & 16.7 & 101 & 23.4 &  & y\\
rainbow & 5 & 21.0 & 125 & 27.5 &  & y\\
fds & 2 & 37.0 & 126 & 29.3 &  & y\\
fda & 78 & 33.9 & 46 & 22.7 & y & y\\
devtools & 80 & 47.0 & 15 & 36.3 & y & y\\
\addlinespace
tidyverse & 89 & 48.4 & 5 & 38.0 & y & \\
FactoMineR & 52 & 41.2 & 37 & 18.5 & y & \\
rms & 54 & 36.8 & 30 & 30.3 & y & y\\
minfi & 38 & 62.4 & 22 & 35.2 & y & y\\
RcmdrMisc & 7 & 19.3 & 43 & 34.6 &  & y\\
\addlinespace
Gviz & 37 & 43.5 & 9 & 35.7 & y & y\\
Rcmdr & 45 & 101.2 & 1 & 81.0 & y & \\
Seurat & 38 & 85.3 & 2 & 54.0 & y & \\
survminer & 27 & 58.2 & 9 & 60.0 & y & y\\
WGCNA & 33 & 52.3 & 2 & 24.5 & y & \\
\addlinespace
weights & 12 & 30.6 & 21 & 35.4 &  & y\\
\bottomrule
\end{tabular}
\end{table*}

Expectedly, many of the top packages listed in Table \ref{tab:table2}
are popular and widely applied in various fields such as general
statistical modeling (\emph{car}, \emph{caret}, \emph{fda}, \emph{rms}
and \emph{FactoMineR}), specific analysis approaches on biological data
(\emph{WGCNA} and \emph{Seurat}), data visualization (\emph{ggpubr},
\emph{Gviz} and \emph{survminer}) and infrastructure-related
applications (\emph{devtools} and \emph{tidyverse}), thus they are depended-on by a great number of child packages. However, developers who depend on
these top packages should be aware of the potential risks that they
introduce.

\subsubsection{How do package dependencies transmit to
children?}\label{how-do-package-dependencies-transmit-to-children}

Package \emph{P} may have multiple child packages. We explored the
distribution of individual heaviness values of \emph{P} on all its child
packages. In Figure \ref{fig:fig_09}B, when packages have high adjusted
HC values, i.e., ranked as top packages affecting their child packages,
Gini indices of the heaviness values get close to 0.2. It indicates
that top heavy packages contribute dependencies almost evenly to their
child packages.

\vspace*{3mm}
\begin{tcolorbox}[colbacktitle=black!20!white,colback=white,coltitle=black,boxrule=0.5pt,sharpish corners,title=\textit{Answer to RQ3},left=1mm,right=1mm,top=1mm,bottom=1mm]

In general, the average unique dependencies that a package contributes
to all its child packages decrease when it has more children. HC metric
is important especially for developers because it also measures the
expected number of additional dependencies if \emph{P} is added as a new
strong parent of their packages. We found many of the top packages with
extremely high HC are already popular in use and we suggest
developers pay more attention to these packages if they want to add them
as new parents of their packages.
\end{tcolorbox}

\subsection{RQ4: heaviness on indirect downstream
packages}\label{heaviness_on_indirect}

HC is a metric of local relation, i.e., the direct parent-child
relation. We next studied how dependency heaviness is uniquely
transmitted to remote downstream of the global dependency graph.

\subsubsection{Remove child packages from
downstream}\label{remove-children-from-downstream}

We observed that, for top packages with the highest heaviness on
downstream packages (abbreviated as HD), they tend to also have high HC
values. In the left panel of Figure \ref{fig:fig_09}C, 478 packages are
common in the top 500 packages with the highest HC values and the top
500 packages with the highest HD values. This implies, for these 478
packages, the downstream packages are mainly composed of child packages.
For this reason, to study the long-range dependency transmission, we
removed child packages from downstream packages, i.e., we only studied
the heaviness on indirect downstream packages (abbreviated as HID). In
the right panel of Figure \ref{fig:fig_09}C, now there are only 129
packages in common in the top 500 packages by HC and by HID, and these 129
packages contribute strong heaviness to both their direct child packages
and remote downstream packages.

\subsubsection{Compare packages with top HC and
HID}\label{compare-packages-with-top-hc-and-hid}

\begin{figure*}[!ht]
{\centering \includegraphics[width=0.95\linewidth]{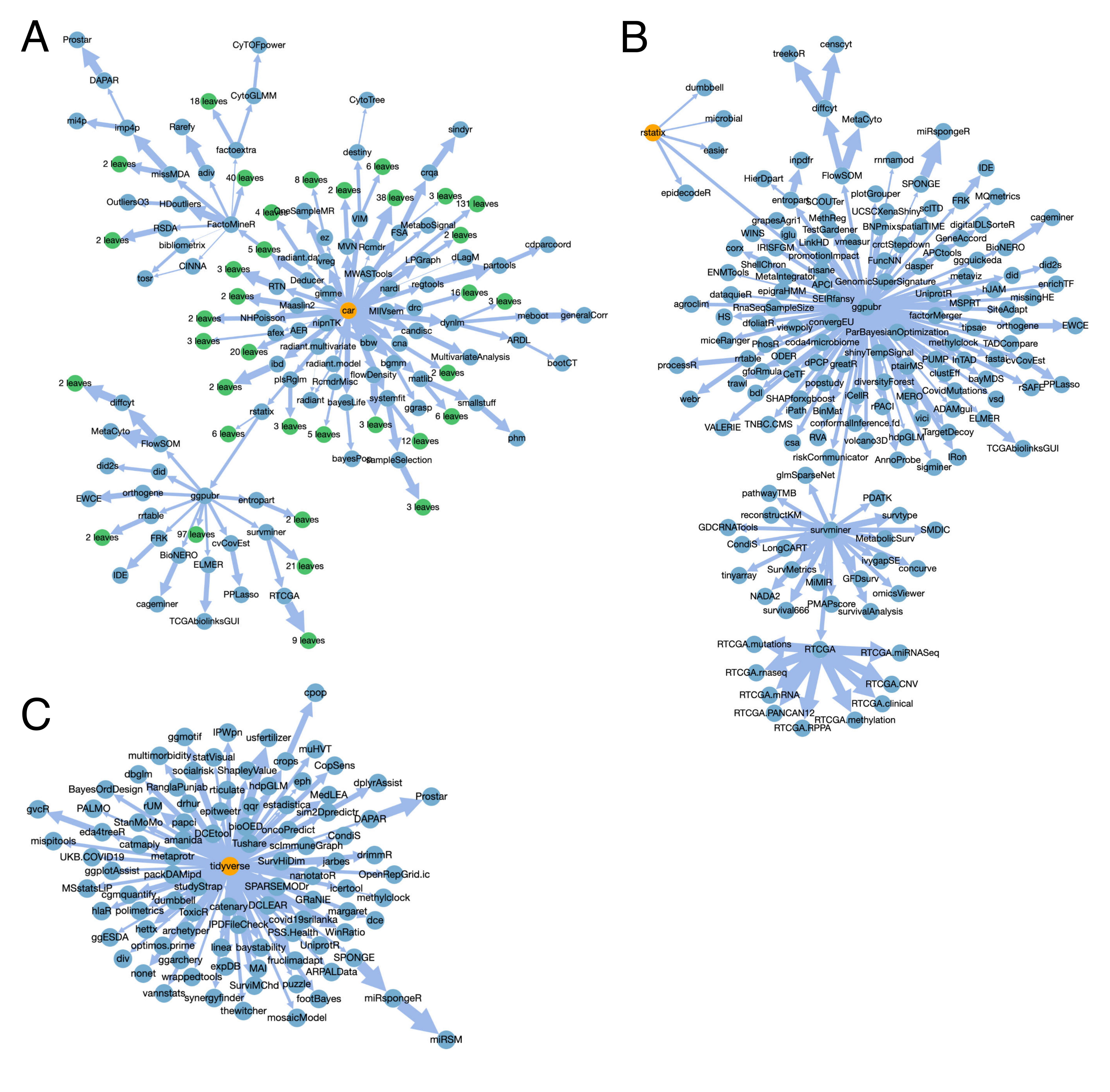}}
\vspace*{-5mm}
\caption{Three examples of dependency graphs of how packages transmit dependencies heaviness to their downstream packages. A) The package \textit{car}. B) The package \textit{rstatix}. C) The package \textit{tidyverse}. The three packages are colored in orange in the graphs. In Figure A, to reduce the size of the graph for visualization, leaf nodes sharing the same parent are grouped and colored in green. Note the three graphs are directed where \textit{car}, \textit{rstatix} and \textit{tidyverse} are the root nodes.}\label{fig:fig_10}
\end{figure*}

Figure \ref{fig:fig_09}D illustrates the distributions of HID for 2,396
(10.9\%) packages with at least one indirect downstream package.
Similarly, distributions also have long tails, which indicates only a
small fraction of packages spread large dependencies to the ecosystem in
the long ranges. In Figure \ref{fig:fig_09}D, we highlighted packages
with adjusted HID \(\ge\) 20. Compared to Figure \ref{fig:fig_06}A, the
list of top packages changes. Table \ref{tab:table3} lists top packages
with adjusted HC \(\ge\) 30 or adjusted HID \(\ge\) 20. To reduce the
length of the table, we additionally set the numbers of downstream
packages (i.e., numbers of child packages + numbers of indirect
downstream packages) \(\ge\) 30. In the 21 packages listed in Table
\ref{tab:table3}, they can be summarized into three categories: i)
having both top HC and HID values (9 packages); ii) only having top HID
values (7 packages); iii) only having top HC values (5 packages). In the
first category, packages not only affect their direct child packages but
also affect indirect downstream packages. An example is the package
\emph{car} which directly affects 183 child packages (with HC of 40.6)
and additionally 435 indirect downstream packages (with HID of 31.9,
Figure \ref{fig:fig_10}A). In the second category, packages have very
few direct child packages, but child packages play as hub packages to
transmit the heaviness to a great number of indirect downstream
packages. One typical example is the package \emph{rstatix}, which only
has 10 child packages, but affects 203 indirect downstream packages.
Figure \ref{fig:fig_10}B demonstrates the heaviness is mainly
transmitted from \emph{rstatix} via a hub child package \emph{ggpubr}
which spreads the dependency flow to 96.2\% of downstream packages of
\emph{rstatix}. In the third category, packages mainly contribute
heaviness only till their child packages and majority of the transmission do
not go further deep in the dependency graph, e.g., the package
\emph{tidyverse} (Figure \ref{fig:fig_10}C). The
downstream dependency graph for any package on CRAN/Bioconductor can be
accessed in the heaviness database in \emph{pkgndep} which will be introduced in Section
\ref{database}.

Similar to HC, CRAN packages have more indirect downstream packages 
than Bioconductor (256.8 vs 136.5, Table
\ref{tab:table1}), but CRAN packages on average transmit less dependency
heaviness to their indirect downstream packages (4.4 vs 8.3, Table
\ref{tab:table1})\footnote{Statistics are based on packages with at least one indirect package.}. Nevertheless, there are more packages with top HID
(adjusted HID \(\ge\) 20) on CRAN than on Bioconductor (25 vs 7).

\vspace*{3mm}
\begin{tcolorbox}[colbacktitle=black!20!white,colback=white,coltitle=black,boxrule=0.5pt,sharpish corners,title=\textit{Answer to RQ4},left=1mm,right=1mm,top=1mm,bottom=1mm]

Only a small fraction of packages spread large dependencies in the
ecosystem in the long ranges. There are three modes of the dependency
transmission: i) Dependency heaviness
is broadly transmitted to both direct child packages and remote
downstream packages; ii) Dependency heaviness is transmitted to remote
downstream via hub child packages; iii) Dependency heaviness is mainly
transmitted locally where their child packages are the ends of the
dependency transmission.
\end{tcolorbox}

\begin{figure*}[!ht]
{\centering \includegraphics[width=0.95\linewidth]{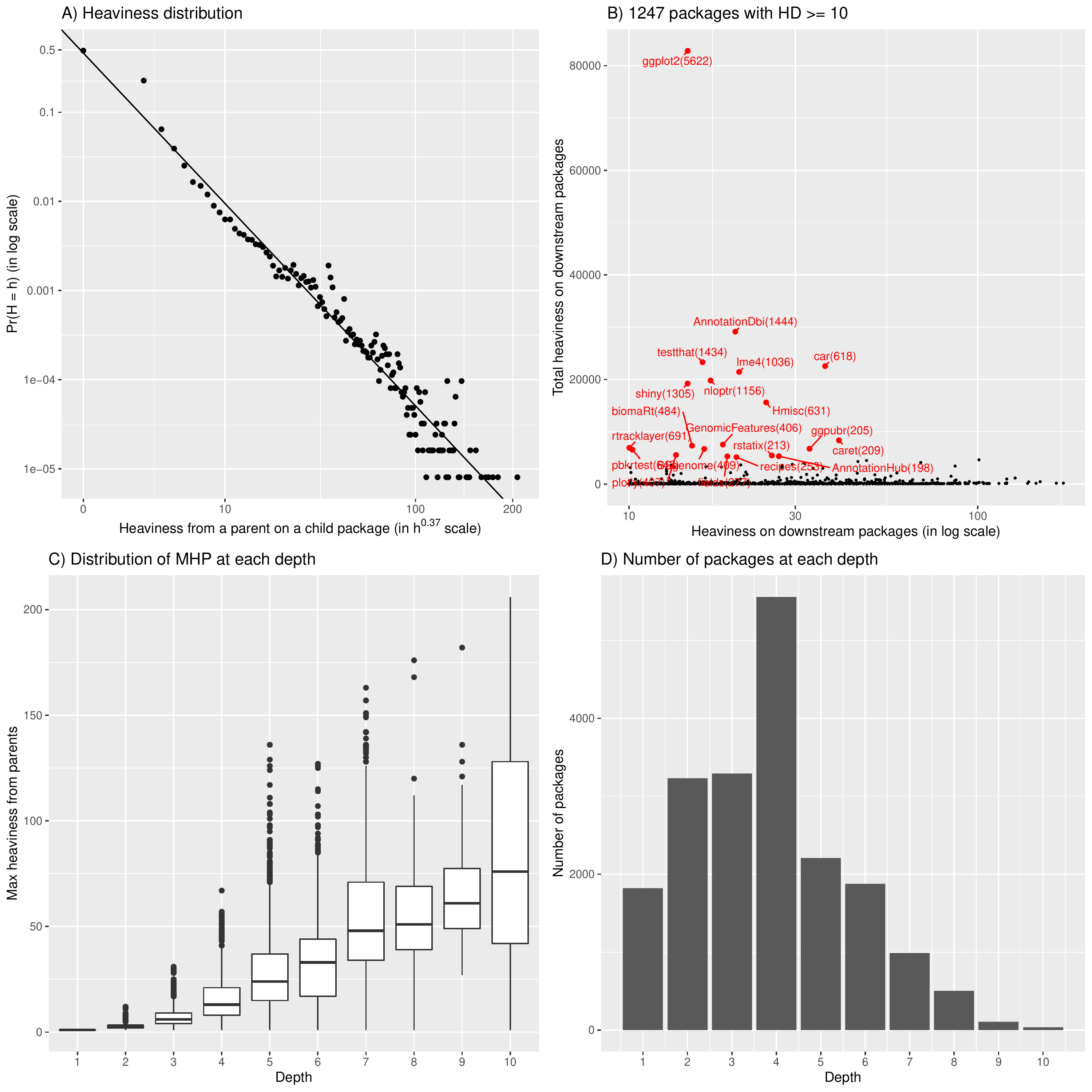}}
\caption{Heaviness analysis on the global dependency graph. A) Distribution of heaviness from a parent on a child package. The line is the fit from a stretched exponential distribution ($R^2 = 0.952$). B) Total heaviness on downstream packages for the packages with HD $\ge$ 10. Packages in red have total heaviness $\ge$ 5,000. Values in the parentheses are the numbers of downstream packages. C) MHP distribution of packages at each depth in the dependency graph. D) Number of packages at each depth. In Figure C and D, packages with depth = 0 (709 packages, i.e., packages with no dependencies) and depth $\ge$ 11 (1 package) are removed for the plot. HD: heaviness on downstream packages; MHP: max heaviness from parents.}\label{fig:fig_11}
\end{figure*}

\subsection{RQ5: Dependency graph analysis}\label{cna_analysis}

In previous sections, the analysis was focused on different heaviness
metrics on individual packages. In this section, we studied the
dependency relations from the aspect of the global dependency graph by
applying the complex network analysis approaches on it.

\subsubsection{The general patterns of heaviness spreads in the
dependency graph}\label{the-general-patterns-of-heaviness-spreads-in-the-dependency-graph}

Denote the global dependency graph as \(G = (V, E)\), where \(V\) is the
set of all packages in the CRAN/Bioconductor ecosystem, \(E\) is the set
of strong dependency relations, and \(G\) is directed, then the
heaviness from a parent on a child is a score associated to an edge in
the graph. Figure \ref{fig:fig_11}A illustrates the distribution of
heaviness which can be approximated as a stretched exponential
distribution \citep{Elton} fitted as

\begin{equation}
  \label{eq:17}
\mathrm{\Pr}(H=h)=0.46\cdot\mathrm{\exp}(-1.66h^{0.37})
\end{equation}

\noindent where \(H\) is the random variable of the heaviness. The model
implies in the ecosystem, there are only a very small amount of
dependency transmissions that are heavy from parent to child packages.
The 95\textsuperscript{th} percentile of all heaviness values is 20.

\subsubsection{Total dependency heaviness to the
ecosystem}\label{total-dependency-heaviness-to-the-ecosystem}

In Section \ref{heaviness_on_children} and \ref{heaviness_on_indirect},
we studied the average heaviness on child packages and indirect
downstream packages. Here we looked at the total heaviness that a
package contributes to the whole ecosystem, i.e., to all its downstream
packages. Using the same denotation in Equation \ref{eq:5}, the total
heaviness of package \emph{P} on all its downstream packages is
calculated as \(h_\mathrm{d} \cdot K_\mathrm{d}\). The total heaviness
also measures the number of reduced dependencies in the whole ecosystem
if \emph{P} is removed. A package with small HD may have a large total
effect simply because it has a huge number of downstream packages. For
example, the package \emph{stringr} only has a small HD of 2.3, but it
affects 5,276 downstream packages, which makes it a top package
contributing dependencies in the ecosystem. Therefore, we only looked at
packages with HD \(\ge\) 10 (1,247 packages left). Figure
\ref{fig:fig_11}B illustrates \emph{ggplot2} is the most influential
package that affects 5,622 downstream packages with a HD value of 14.7.
It uniquely contributes in total 82,830 dependencies to the
ecosystem. Another example is the package \emph{car} which affects 618
downstream packages but with a higher HD value of 35.6, uniquely contributing in
total 20,001 dependencies. All of the top 20 packages (filtered
by total heaviness on downstream \(\ge\) 5,000) listed in Figure
\ref{fig:fig_11}B affect a large number of downstream packages, with an
average number of 882. The top 20 packages (0.1\% of all packages)
contribute 18.9\% unique dependencies to the whole ecosystem. This
suggests if developers of these top packages can manage to reduce
dependencies of their packages, it will greatly reduce the risks they bring to the whole
ecosystem.

\subsubsection{Depth of the dependency heaviness
transmission}\label{depth_from_upstream}

In Section \ref{heaviness_from_parents}, we observed dependency
heaviness on top packages with the highest MHP are accumulated from
very remote upstream. Here we studied dependency heaviness
transmission for all packages in the ecosystem. For each package, we
calculated its depth in the graph as the maximal distance from all its
upstream packages where the distance between two packages is the length
of the shortest path in the directed graph. Figure \ref{fig:fig_11}C
illustrates the distribution of MHP at each depth, and Figure
\ref{fig:fig_11}D illustrates the number of packages at each depth. The
two figures clearly show, when a package locates more downstream in the
graph, it receives larger dependency heaviness from its parent.

\begin{figure*}[!ht]
{\centering \includegraphics[width=0.95\linewidth]{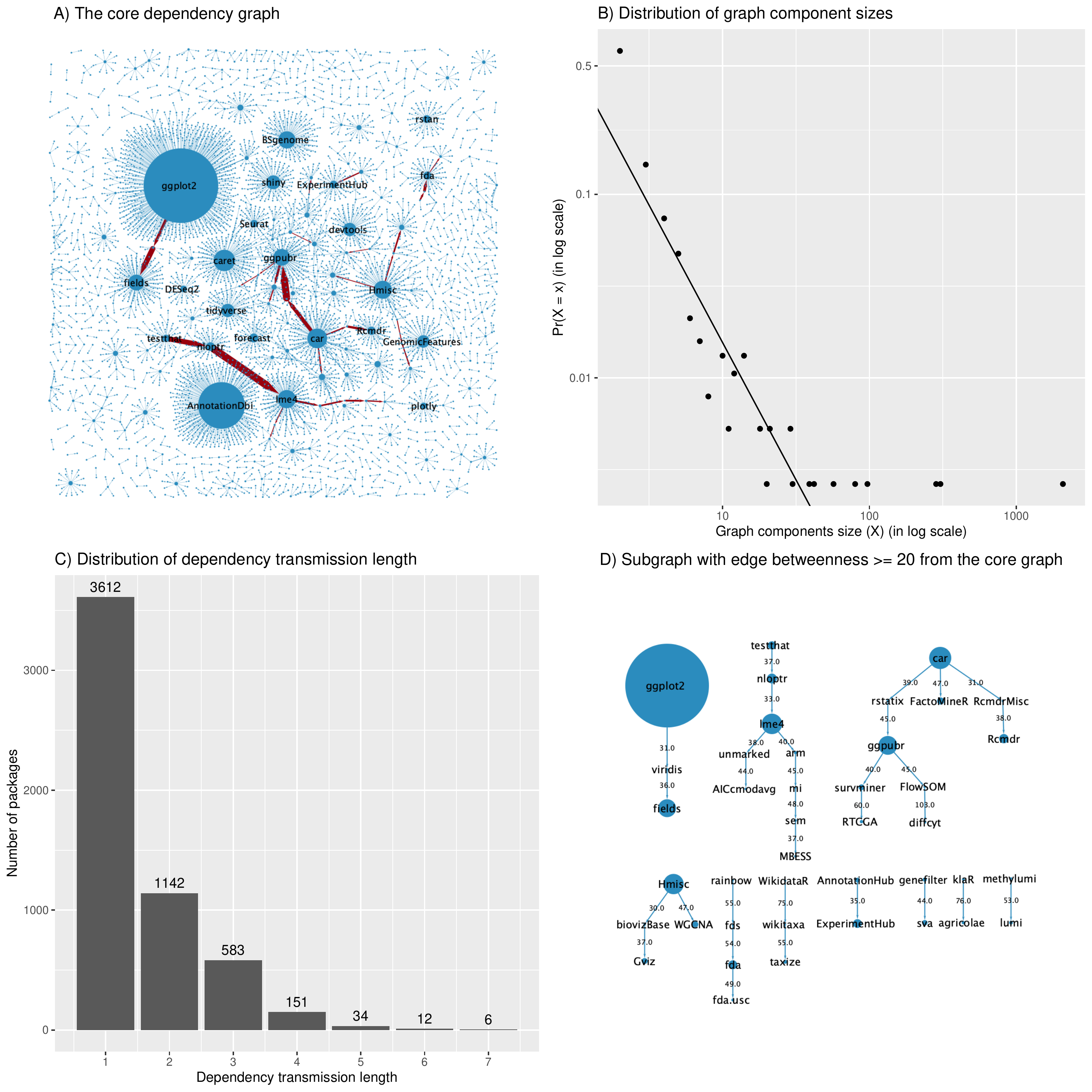} }
\caption{Analysis on the core dependency graph. A) Network visualization of the core graph. Node size is mapped to out-degree, i.e., number of child packages. Labels of hub packages with out-degree $\ge$ 30 are added. Edge width is mapped to the betweenness and edges with betweenness $\ge$ 20 are highlighted in red. B) Distribution of graph component sizes. The line is the fit of the power-law distribution ($R^2 = 0.845$). The largest five data points are removed from fitting. C) Distribution of dependency transmission length of packages in the core graph. D) The subgraph that only contains high betweenness edges from Figure A. Node size is mapped to the out-degree calculated from the core graph. Values on edges are the heaviness from parents to corresponding child packages.}\label{fig:fig_12}
\end{figure*}

\subsubsection{Core graph}\label{core_graph}

To study how the dependency heaviness is transmitted in the global
dependency graph, we constructed a core graph where heaviness from a
parent on a child is \(\ge\) 30. The core graph includes 4,302 packages
(19.5\%), 3,950 strong dependency relations (3.2\%) and 44.2\% heaviness
flows of the complete graph (measured as the fraction of total heaviness
in the core graph and in the global graph).

The core graph is visualized in Figure \ref{fig:fig_12}A. It can be
easily observed that there are several hub packages that transmit large
heaviness to their direct child packages. The top two packages
\emph{ggplot2} and \emph{AnnotationDbi} contribute heavy dependencies to
their 442 and 273 child packages directly. We also observed there are a
huge number of isolated and small graph components (i.e., maximally
connected subgraphs) where heaviness is only transmitted locally. Out of
the total 379 graph components, 352 components only have size \(\le\)
10; however, the other 27 components include 77.0\% of packages in the
core graph. The largest component leaded by \emph{ggplot2} includes
2,082 packages (48.4\% of all packages in the core graph). The size of
the graph component approximately follows a power-law distribution
(Figure \ref{fig:fig_12}B).

We next studied how deep a package \emph{P} can transmit heavy
dependencies to downstream of the core graph. For this, we defined a
metric named ``dependency transmission length''. Assume \emph{P} is
reachable to \(N_\mathrm{leaf}\) leaf packages\footnote{Package \emph{A} is reachable to package \emph{B} in the dependency graph when the distance from \emph{A} to \emph{B} is finite.} where a leaf package
has out-degree of zero in the graph, let \(d_i\) be the distance from
\emph{P} to the \textit{i}\textsuperscript{th} leaf package which is
the length of the shortest path from \emph{P} to the leaf package,
then the dependency transmission length denoted as \(l\) for \emph{P} is
calculated as

\begin{equation}
  \label{eq:18}
l = \max_{i \in \{1..N_\mathrm{leaf}\}} d_i .
\end{equation}

Figure \ref{fig:fig_12}C illustrates the distribution of dependency
transmission length. It shows the majority of the transmission from
\emph{P} only has a length of 1, which can be easily confirmed by Figure
\ref{fig:fig_12}A. But there also exist long paths where heavy
dependencies can be continually transmitted to \emph{P}'s deep downstream packages
(examples in Figure \ref{fig:fig_12}D). Note the depth analysis here is
different from that in Section \ref{depth_from_upstream}. In Section
\ref{depth_from_upstream}, we studied how the dependency heaviness
\emph{P} inherits from its upstream, while in this section we studied how the
dependency heaviness is transmitted to \emph{P}'s downstream.

\begin{figure*}[!htp]
{\centering \includegraphics[width=\linewidth]{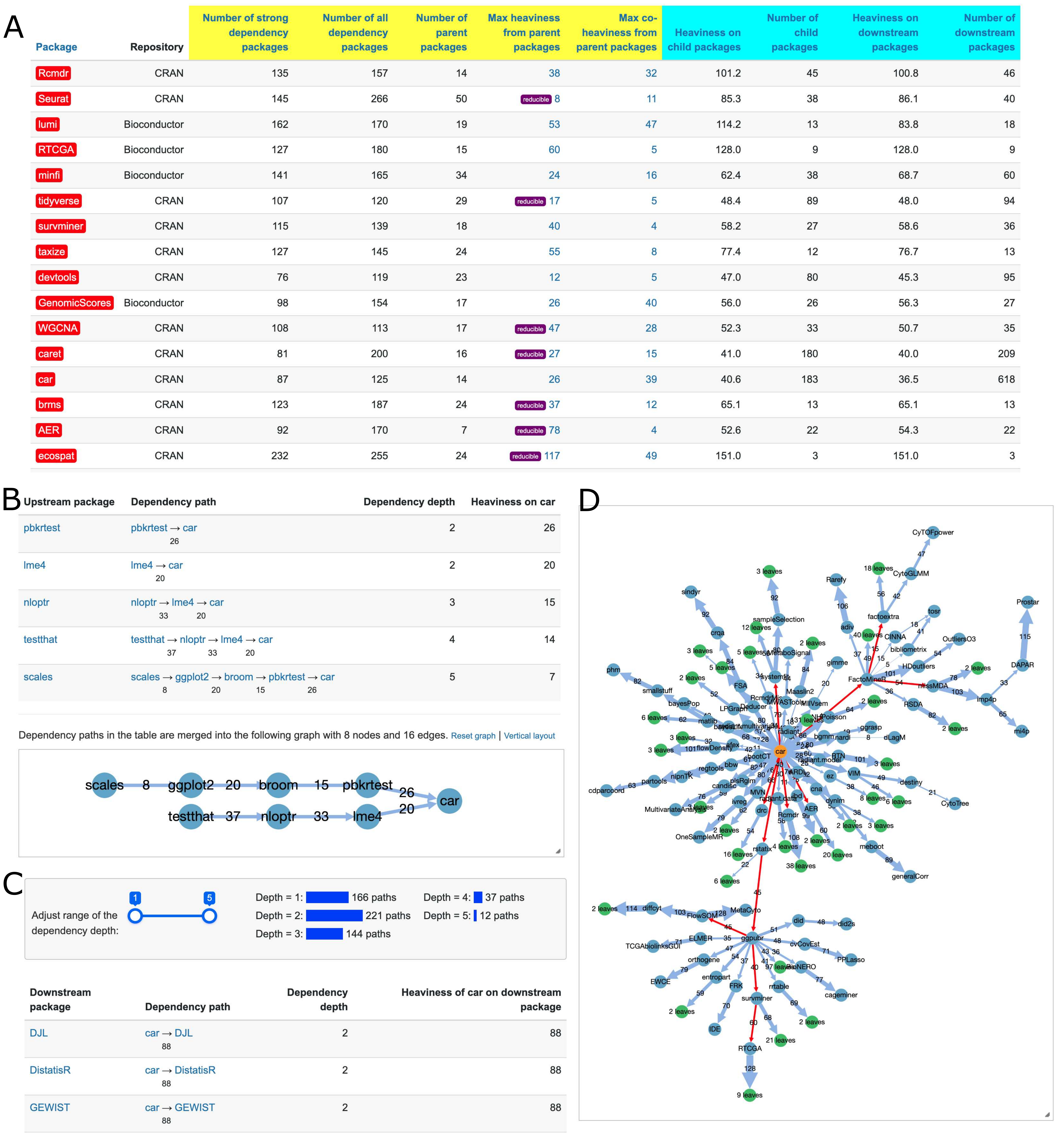}}
\caption{Web-based database of the dependency heaviness analysis for all R packages on CRAN/Bioconductor. A) The global heaviness table of all packages. B) Upstream of a package. Dependency path from every upstream package is listed in a table and visualized in a graph. C) Downstream of a package. Dependency path to every downstream package is listed in a table. The table can be filtered by selecting a range of depths of the dependency paths. D) A graph of the downstream dependency graph. Green nodes represent groups of leaf packages that connect to the same parent package. Edges with high betweenness are highlighted in red. The cutoff of betweenness is selected as the `elbow' of the curve of sorted betweenness of all edges in the graph. Values on edges are the heaviness from parents to corresponding child packages.}\label{fig:fig_13}
\end{figure*}

\subsubsection{Key dependency paths}\label{key-dependency-paths}

The betweenness of an edge measures the number of shortest paths in the
graph that pass through the edge. In the context of the dependency
graph, the edge betweenness measures the amount of heaviness flows that are
transmitted via a parent-child pair. In Figure \ref{fig:fig_12}A, edges
with betweenness \(\ge\) 20 are highlighted in red and a subgraph that
only contains these high betweenness edges is induced in Figure
\ref{fig:fig_12}D. We named the subgraph as ``key dependency paths''.
The key paths transmit 25.2\% of the total heaviness flows (measured as
the fraction of total betweenness in the key paths and in the core
graph) while only including 0.8\% edges and 1.0\% packages from the core
graph. In most cases, in the key paths, the heaviness is transmitted
from hub packages, however there are exceptions. For example,
\emph{rstatix} only has a few child packages (10 in the global graph and
4 in the core graph), but it connects two hub packages \emph{car} and
\emph{ggpubr} as a bridge to continue the dependency transmission
(Figure \ref{fig:fig_12}D). Another similar example is \emph{viridis}
that connects \emph{ggplot2} and \emph{fields} (Figure
\ref{fig:fig_12}D). Besides that, we also found there are long paths
such as from \emph{testthat} to \emph{MBESS} with length of 6, and from
\emph{car} to \emph{diffcyt} or \emph{RTCGA} with length of 5, where the
heavy dependencies can be continually transmitted through.

\vspace*{3mm}
\begin{tcolorbox}[colbacktitle=black!20!white,colback=white,coltitle=black,boxrule=0.5pt,sharpish corners,title=\textit{Answer to RQ5},left=1mm,right=1mm,top=1mm,bottom=1mm]

We constructed a core graph which transmits heavy dependencies in the
ecosystem. We revealed graph components and key paths that transmit
major heaviness in the ecosystem. We found that hub packages mainly
transmit dependency heaviness only to their child packages, thus
locally. When a package locates more deep in the downstream of the
dependency graph, it preferably inherits larger dependency heaviness
from the ecosystem.
\end{tcolorbox}

\section{The open database}\label{database}

We have integrated the dependency heaviness analysis for the
CRAN/Bioconductor ecosystem in the package \emph{pkgndep} as a web-based
database (We call it the heaviness database, Figure \ref{fig:fig_13}). It provides detailed analysis
reports on the dependency heaviness both from direct parent-child
relations and remote upstream-downstream relations. The heaviness
database can be simply accessed with the function
\texttt{dependency\_database()} from the \emph{pkgndep} package.

The database has two parts: a summary table of all packages and analyses of
individual packages. Figure \ref{fig:fig_13}A illustrates the global
table of the heaviness analysis of all packages. Columns in the table
are separated into two groups highlighted in yellow and blue, which
correspond to metrics from upstream (e.g., MHP) and on downstream (e.g.,
HC, HD and HID). Packages with adjusted HC \(\ge\) 30 are highlighted in
red to emphasize that they have high impacts on the ecosystem. If a
package only imports a limited number of functions from the heaviest
parent, the package is marked as `reducible' in purple, which implies
possibility to reduce its upstream dependencies for developers (see our
suggestions in Section \ref{parent_developer}).

The database contains comprehensive tools for querying dependencies for
individual packages. For a package \emph{P}, the database allows to
explore how the dependencies are inherited from parent or upstream to
\emph{P}, and how the dependencies are transmitted from \emph{P} to its
child or downstream. In the direct parent and child dependency results,
there are tables showing heaviness-related metrics. It also lists the
``Imports'' information (i.e., how classes and methods are imported from
parents  to \emph{P} or from \emph{P} to its children)
which is automatically parsed from the NAMESPACE files of corresponding
packages. In the upstream dependency results, the dependency path from
each upstream package to \emph{P} is listed, which is the shortest path
from the upstream package to \emph{P} in the global dependency graph
(Figure \ref{fig:fig_13}B). There is also an interactive graph that
shows how the heaviness is accumulated (Figure \ref{fig:fig_13}B).
Similarly, in the downstream dependency results, the dependency path
from \emph{P} to every downstream package is listed, which can be
further filtered by the depth to the downstream packages (Figure
\ref{fig:fig_13}C). The downstream dependency graph is also visualized
as an interactive graph (Figure \ref{fig:fig_13}D). Normally, the
downstream dependency graph is large. In order to reduce the graph size
for visualization, leaf packages are grouped into a single node if they
have the same parent (Figure \ref{fig:fig_13}D). Additionally, edges
with high betweenness in the downstream dependency graph are highlighted
in red, which correspond to the key paths that transmit major dependency
heaviness from \emph{P}.

\section{Considerations for
developers}\label{consideration_for_developers}

The heaviness analysis is especially useful for developers. As this study was motivated and accumulated from the author's experience as an active developer of R packages\footnote{\url{https://jokergoo.github.io/software/}.}, we proposed the following considerations from three different aspects from a developer's perspective. Note the three aspects are associated while not
isolated. In the following subsections, we discussed each aspect with
several examples to demonstrate how heaviness analysis benefits
developers. The dependency heaviness analysis for all example packages
mentioned in this section can be accessed in the heaviness database in
\emph{pkgndep}.

\subsection{How to properly handle the dependency of a
package?}\label{parent_developer}

It is a good practice to keep package dependency as simple as possible.
However, there is always a balance between the compactness of
dependencies and the comprehensiveness of a package's functionalities.
If a package has a parent showing high heaviness (e.g., with high MHP),
it is a sign that reduction of the dependency complexity should be considered.
We have the following three suggestions that developers may consider.

First, if package \emph{P} only imports one or a small amount of simple
functions from its parent \emph{A}, heavy dependencies from \emph{A} can
be avoided by directly implementing functions with the same functionalities
as the original ones. For example, as we have demonstrated in our
previous study \citep{pkgndep}, the package \emph{mapstats} has
a heavy parent \emph{Hmisc} with heaviness of 49 where only a single
function \texttt{capitalize()} is imported to \emph{mapstats}.
\texttt{capitalize()} is an extremely simple function that only
capitalizes the first letter of a word. It can be easily re-implemented
by developer's own to get rid of the 49 unnecessary dependencies.

Second, on CRAN/Bioconductor, it is common that there are several
packages providing the same functionalities for an analysis task. Then
if \emph{P} depends on a heavy parent, the developer can look for a
light dependency package which provides the same functionality
as the heavy one. For example, the package \emph{biovizBase} has a heavy
parent \emph{Hmisc} with heaviness of 30 where a single function
\texttt{bezier()} is imported to \emph{biovizBase}. \texttt{bezier()} is
for generating Bézier curves and the use of \emph{Hmisc} can be replaced
with other lighter packages that also generate Bézier curves, e.g., a package
called \emph{bezier} but with zero additional dependency. As a note,
this optimization on \emph{biovizBase} is even more meaningful because
the reduction of the 30 extra dependencies on \emph{biovizBase} can
additionally reduce the dependencies of its 631 downstream packages with
an average reduced dependencies of 20 for each.

Third, Some packages aim to be a ``toolkit'' to provide comprehensive
analysis by integrating many other packages. This increasing
comprehensiveness also brings the expansion of dependencies. For
example, the package \emph{singleCellTK} provides comprehensive tools
for analyzing single-cell RNASeq data by depending on 82 parents and in total 369 strong
dependencies\footnote{\textit{singleCellTK} is the package with the largest number of parents and the second largest number of strong dependencies on CRAN/Bioconductor.}.
This makes it very vulnerable to failures from upstream packages and it
is not friendly for users to install. Nevertheless, for such toolkit
packages, there are always core functionalities that are more frequently
used by users and optional functionalities that are less used. The huge
dependencies of such packages can be reduced by moving parents that only
provide optional functionalities to weak parents, then it dramatically
reduces the total strong dependencies (see methods in Section \ref{flexible-control-of-dependencies}). Take the package \emph{cola}
which is also developed by the author as an example. \emph{cola}
provides consensus clustering analysis as its core functionality, and it
also provides comprehensive functions for downstream analysis. According
to our experience, we found some downstream analyses that depend on
heavy parents are rarely used by users, thus, we set them as weak
parents. With this strategy, the strong dependency of \emph{cola} was dramatically
reduced from 252 to 64.

Nevertheless, there are also packages with no parent contributing
significantly high dependency heaviness. In this case, optimizing the
dependencies is difficult. An example is the package \emph{Seurat} which
has 50 parents and 145 strong dependencies, but the heaviest parent only
contributes 8 additional dependencies (MHP = 8), thus optimization on
one or only a few parents won't dramatically reduce dependencies of
\emph{Seurat}. Besides that, there are also scenarios where reduction of
heavy parents could not be performed: (i) A heavy parent provides core
functionality to its child package; (ii) S4 methods or S4
classes\footnote{S4 is an object oriented system in R.} are imported
from a parent package; (iii) A child package depends on the C/C++
headers from a
parent\footnote{Then the parent must be put in the ``LinkingTo'' field of the child package.}.

\subsection{How is dependency heaviness accumulated to a
package from upstream?}\label{dev_upstream}

\begin{figure}[t]
{\centering \includegraphics[width=1\linewidth]{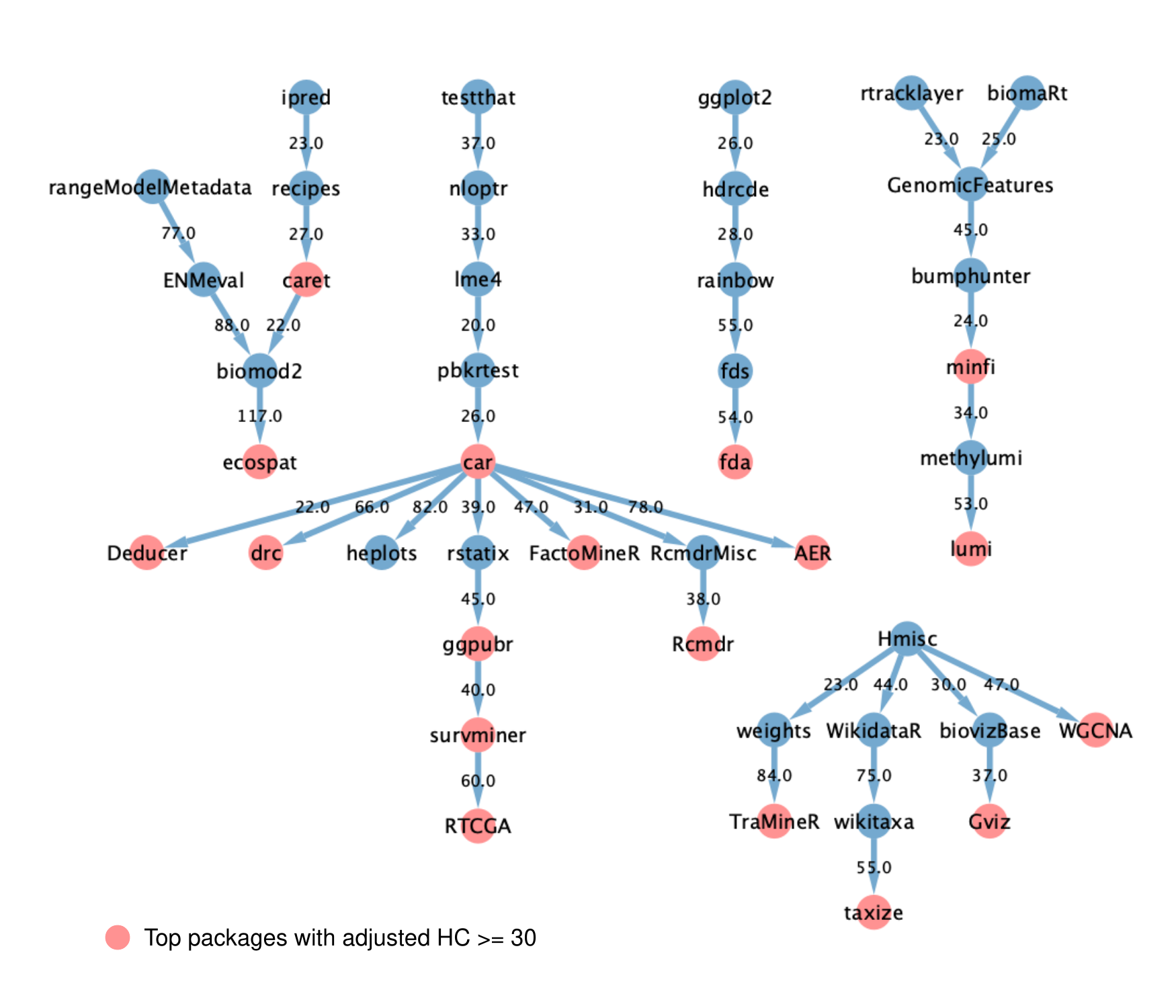}}
\vspace{-10mm}
\caption{Upstream of packages with adjusted HC $\ge$ 30. For simplicity, only edges with heaviness $\ge$ 20 and graph components with size $\ge$ 5 are kept in the figure. Values on edges are the heaviness values from parents to corresponding child packages. HC: heaviness on child packages.}\label{fig:fig_14}
\end{figure}

Dependencies from remote upstream cannot be directly controlled by the
developer, but it is still useful for understanding how the heaviness is
accumulated to his package. Here we take top packages with the highest
adjusted HC (Section \ref{heaviness_on_children}) as examples, because
they also inherit huge dependency heaviness from their upstream and they
play important roles in the ecosystem. Figure \ref{fig:fig_14}
illustrates upstream packages that transmit major dependency heaviness to
top HC packages. In general, dependencies are transmitted in
the long ranges, which agrees with the results in Section
\ref{heaviness_from_parents}. We observed there are three different
modes of dependency accumulation.

\begin{enumerate}
\def\labelenumi{\arabic{enumi}.}
\item
  \textbf{Heaviness is accumulated from multiple heavy parents.} For
  example, package \emph{ecospat} inherits unique dependencies from
  \emph{ENMeval} (with a heaviness of 74 uniquely transmitted to \emph{ecospat}) and \emph{caret} (with a heaviness of
  19 uniquely transmitted to \emph{ecospat}) separately.
\item
  \textbf{A group of packages inherit heavy dependencies all from the
  same heavy upstream package.} For example, package \emph{TraMineR},
  \emph{taxize}, \emph{Gviz} and \emph{WGCNA} all inherit huge unique
  dependencies from the same upstream package \emph{Hmisc}.
\item
  \textbf{The heaviness transmission can be traced back to a remote
  upstream package.} As a typical example, dependency transmission to package
  \emph{RTCGA} can be traced back to the package \emph{car} which
  additionally affects several other packages. If tracing further to
  \emph{car}'s upstream, we can see the heavy dependencies are actually
  from the package \emph{testthat}.
\end{enumerate}

Once we have revealed the upstream source of the heaviness, this leads
us to the question which is asked in the next section: how does the
dependency heaviness spread in the ecosystem from the ``source''? In
particular, we will explore the impacts of \emph{testthat} and
\emph{Hmisc} on their downstream packages.

\subsection{How to reduce the dependency heaviness spreading to the
downstream?}\label{how-to-reduce-the-dependency-heaviness-spreading-to-the-downstream}

When a package becomes popular in the ecosystem, it is very important
for the developer to carefully manage its dependency size. If the
developer introduces new features that need extra dependencies, he
should be aware of the risks that are also brought to the ecosystem. We
suggest the developer also paying attention to the current set of
parents of his package, and trying to compact its dependency size.

According to our observation, there are still a large number of packages
on CRAN/Bioconductor whose dependency sizes can be reduced. Here we take
the package \emph{AnnotationDbi} as an example. \emph{AnnotationDbi} is
a hub package that contributes dependency heaviness to 1,444 downstream
packages in the ecosystem (with HD of 20.2, mainly on Bioconductor). It was also mentioned as a
high impact package in Section \ref{co_heaviness_pair} and Section
\ref{core_graph}. By exploring how \emph{AnnotationDbi}'s parents
contribute their functionalities to \emph{AnnotationDbi}, we found a
package \emph{KEGGREST} contributes 20 unique dependencies to
\emph{AnnotationDbi} and these extra dependencies are transmitted to all
downstream packages of \emph{AnnotationDbi}. A close inspection shows
only one function \texttt{KeggList()} is imported to
\emph{AnniotaitonDbi} and this functionality is very rarely used by its
massive downstream packages. If \emph{KEGGREST} can be set as a weak
parent of \emph{AnnotationDbi}, on average 9.5 dependencies can be
reduced for every of its 1,444 downstream packages. This actually
implies even a small reduction of dependencies of a hub package will
have a huge impact on the whole ecosystem.

It is also interesting to explore how dependencies from hub packages are
transmitted to downstream packages in the ecosystem. This helps to
reveal the ``problematic sites'' of the dependency transmission and later
developers can propose possible solutions to fix them. We found the
following two typical scenarios where heavy dependencies are improperly
transmitted to the ecosystem and they can actually be avoided.

\textbf{Inefficient use of a parent package.} In Section
\ref{dev_upstream}, we revealed a long dependency heaviness transmission
originated from \emph{testthat}. A closer look shows \emph{testthat}
behaves as a hub package in the ecosystem which has 130 child packages
and 1,304 indirect downstream packages, with HC of 18.2 and HID of 15.8
respectively\footnote{The downstream dependency graph of \textit{testthat} can be accessed in the heaviness database in \textit{pkgndep}.}.
This is quite unexpected because \emph{testthat} is a package mainly for
package unit testing purpose and it is normally put in a package's
``Suggests'' field\footnote{\url{https://r-pkgs.org/testing-basics.html}.}. We found there are two reasons for its high impact
in the ecosystem. First, majority of the heaviness from \emph{testthat}
are transmitted via its child package \emph{nloptr} to downstream. In
total 18,933 extra unique dependencies (81.3\% of all from
\emph{testthat}) are transmitted to its 1,156 (80.6\%) downstream
packages via \emph{nloptr}. \emph{nloptr} declares \emph{testthat} as
its strong parent because \emph{nloptr} performs unit testing on its C++
code which requires a header file from \emph{testthat}, thus
\emph{testthat} must be put in its ``LinkingTo'' field. If by some
means, \emph{nloptr} can get rid of the strong dependency on
\emph{testthat}, e.g., by using another unit testing tool on C++ code,
on average 16.4 dependency heaviness for each of its 1,156 downstream packages
can be reduced. 

Secondly, we also observed \emph{testthat} is used
inefficiently in its 129 other child packages (excluding \emph{nloptr}).
In these packages, developers use \emph{testthat} to perform object
validation in the source code, e.g., to compare whether two objects are
equal (by \texttt{expect\_equal()}) or to validate a text output (e.g.,
by \texttt{expect\_match()}). \emph{testthat} provides comprehensive
tools for unit testing on packages. while it would be too heavy if it is
directly used in package's source code just for simple validation.
Actually, developers can replace \texttt{expect\_*()} functions from
\emph{testthat} with self-implemented code very easily. For example
\texttt{expect\_equal(x,\ y)} can be replaced by \texttt{x\ ==\ y}
(assuming \texttt{x} and \texttt{y} are two scalars), and
\texttt{expect\_match(text,\ regexp)} can be replaced by
\texttt{grepl(regexp,\ text)}. In this way, large amount of extra
dependencies of \emph{testthat} can be reduced.

\textbf{A hub package providing a wide range of functionalities.} If a
hub package provides a wide range of functionalities, it is very likely
that it also inherits a large number of packages from upstream (We have partially discussed it in Section \ref{parent_developer}). All the
dependencies of the hub package are transmitted to the downstream even though child
packages may only import a limited number of functions from it. Here we
take the package \emph{devtools} and \emph{Hmisc} as two examples.
\emph{devtools} provides functionalities for package development, but in
the ecosystem it has 80 child packages with HC of 47 and 15 indirect
downstream with HID of 36.3. A deep inspection shows most of its child
packages import the function \texttt{install\_github()} to install
dependencies that are directly from their development branches on
GitHub. There are two possible optimizations. First, dependency on \emph{devtools} can be set as a weak
parent because \emph{devtools} does not contribute to the
functionalities of its child packages. And second, the
installation functionality in \emph{devtools} can be separated into a
new and light package. Actually, \texttt{install\_github()} has already
been moved to a new package
\emph{remotes}\footnote{The first version of \textit{remotes} was released in 2016.},
and developers can consider to migrate from \emph{devtools} to
\emph{remotes}. 

As a second example, also as we have mentioned in
Section \ref{co_heaviness_pair} and Section \ref{dev_upstream},
\emph{Hmisc} behaves as a heavy hub packages with 248 child packages
(with HC of 29.4) and 383 indirect downstream packages (with HID of 20).
\emph{Hmisc}, as its name tells, provides a huge collection of functions
for \emph{``data analysis, high-level graphics, utility operations,
functions for computing sample size and power, simulation, importing and
annotating datasets, imputing missing values, advanced table making,
variable clustering, character string manipulation, conversion of R
objects to LaTeX and HTML code, and recording
variables''}\footnote{\url{https://CRAN.R-project.org/package=Hmisc}.}. It has
18 parents and in total 67 strong dependencies from upstream. The
dependency heaviness analysis on \emph{Hmisc} reveals \emph{ggplot2},
\emph{viridis} and \emph{htmlTable} contribute majority (59.7\%) of
dependencies to \emph{Hmisc}\footnote{Heaviness analysis of \emph{Hmisc} can be performed in the heaviness database in \emph{pkgndep}.}. The first two are for data visualization
and the last one is for report generation. A deep inspection of how
\emph{Hmisc} is used in its child packages shows the visualization and
reporting are very rarely used. Thus, similar as \emph{devtools}, if
\emph{Hmisc} can separate its visualization and reporting parts out as a
separated package, it can save on average 18.5 extra dependencies for
every of its 631 downstream packages.

\section{Discussion}\label{discussion}

Dependency analysis is an important topic for studying package ecosystems. One
of the aims is to discover top packages that have major impacts on the
dependency transmission in the ecosystem. Number of dependents is a
widely-used metric that measures the local impact of how important a package
is in contributing to other packages’ functionalities \citep{Mora_CNA, Korkmaz2019}. By taking the
ecosystem as a whole, researchers studied the vulnerability of packages to the
failures caused by recursive dependencies from upstream. Then, a more useful
metric, the number of transitive dependencies, is proposed \citep{Mora_CNA, decan_empirical_2019}. Besides the
transitive effects accumulated from upstream, researchers also looked at the
number of transitive dependents to study the indirect influences
on downstream of the ecosystem \citep{decan_empirical_2019, abate}. Although these metrics are
useful for understanding the attributes of the ecosystem, they are used as
descriptive statistics in current studies \citep{decan_empirical_2019}. These analyses are usually global and they are
limited for developers because they provide almost no practical help on how to
manage dependencies of their packages. After high impact packages are
discovered from the ecosystem, naturally there will be the follow-up questions
to ask, such as how the dependencies are transmitted from upstream to the high
impact packages or how the dependencies are transmitted from the high impact
packages to downstream? This may bring more questions such as can we find the
most important part of the dependency transmission in the ecosystem? As a package may have
multiple dependencies, either direct or transitive, it is quite common that
individual dependencies have different levels of influences on the package.
This implies, to understand the ecosystem deeper, we need to shift the focus
from package-centered to dependency relation-centered, i.e., to find which
dependency relation is more important with regard to transmitting
dependencies.  For this purpose, we proposed a new metric named “dependency
heaviness” which quantitatively measures the unique contribution of dependency
from a parent to a child. Dependency heaviness is also based on transitive
dependencies, but it measures from a different aspect. With this metric, we
can easily identify which parents are heavy with regards to how they
contribute dependencies to a child package.

Based on the direct dependency heaviness from a parent to a child, we extended
the heaviness definitions to a broader range to study the patterns of unique
dependency flows in the ecosystem. We aimed to answer the question of how the
dependencies are uniquely transmitted through the ecosystem. We first explored
the heavy dependency inheritance from parent and upstream packages. This
analysis might be less interesting for software engineering because top
packages with this metric are basically special cases. They are preferably located at the end
of the dependency transmission chain and with no dependents, thus having no
large impact on the ecosystem. But as they inherit heavy dependencies from
upstream, it is still interesting to explore how the dependencies are
accumulated. We found the deeper a package is located in the ecosystem, the more
likely it inherits heavier dependencies. 

We next explored how packages
transmit unique dependencies to the downstream of the ecosystem with two
metrics of HC and HID. This analysis is more important because top packages
with the highest HC or HID have major impacts on the dependency transmission in the ecosystem.
HC is a more practical metric because it also measures the expected number of
additional dependencies if a package is included as a new parent of a
developer’s package. HC, although it is also summarized from all its child
packages, can generate different results from current studies. For example,
the package \emph{Rcpp} has 2,795 child packages and it is the package with the third
most child packages in the ecosystem. As a hub package, code breaks
of \emph{Rcpp} will affect a large number of other packages. In this sense, \emph{Rcpp} can be treated
as the source of the “risk” in the ecosystem. However, from the aspect of dependency
heaviness, it only has a HC of 0.58, which means it is an extremely light package.
HC and HID focus more on the role of a package as an intermediate package
receiving dependencies from upstream and transmitting to downstream. So they
are more like bi-directional metrics of the dependency transmission.
Additionally, HC and HID focus more on the influences on individual packages while
not on the whole set of dependents.

We applied network analysis on the dependency graph. Being different from
network analysis in current studies which take the graph as unweighted
\citep{Mora_CNA}, we studied the graph by weighting edges with dependency heaviness. In
this way, we can distinguish which relations are more important in the graph.
Based on the edge weights, we extracted a core graph and key paths that
transmit major dependencies in the ecosystem. A deeper analysis on the
dependency graph revealed heavy dependencies can be accumulated from very far
upstream, and hub packages transmit heavy dependencies most locally.

We have implemented the complete analysis as a web-based database that can be
easily accessed from the \emph{pkgndep} package. The database provides detailed
statistics for various heaviness metrics both from upstream and downstream
packages. It also provides comprehensive dependency analysis for individual
packages. It helps developers understand how the dependencies are accumulated
to their packages from the ecosystem and how the dependencies spread to the
downstream of their packages. We have extensively used
it to study the dependency chains of example packages in Section \ref{consideration_for_developers}.

\section{Limitations and future work}\label{future}

In Section \ref{heaviness_from_parents}, we studied patterns of maximal
heaviness from parents (MHP) in the ecosystem. As we have demonstrated,
if a package suffers heavy dependencies from its parents, in many cases,
the heaviest parent dominantly contributes unique dependencies to it.
Nevertheless, there are still cases where there are more than one
dominantly heavy parents. For example, the package
\emph{pathwayTMB}\footnote{\textit{pathwayTMB} inherits 215 strong dependencies from its 16 parent packages.}
has two heaviest parents of \emph{clusterProfiler} and \emph{survminer}
which contribute dependency heaviness of 62 and 55 mutually exclusively.
One reasonable hypothesis is that if package \emph{P} has more than
one heavy parent, these heavy parents are responsible for different
analysis tasks in \emph{P}. Indeed, for the two heavy parents of
\emph{pathwayTMB}, \emph{clusterProfiler} performs gene set enrichment
analysis and \emph{survminer} performs survival analysis, thus they
introduce dependencies from different sources. Nevertheless, such
scenarios where \emph{P} has multiple heavy parents should not be often,
but it provides a complementary view on the ecosystem.

In Section \ref{coheaviness}, we defined the co-heaviness metric from
two parents. We can define co-heaviness from more than two parents in a
similar way. However, as co-heaviness only measures the unique number of
dependencies that a group of parents simultaneously bring in, with
considering more parents, the co-heaviness value will decrease and it
would not be a proper metric for studying the ecosystem.

Our study was only focused on the CRAN and Bioconductor ecosystems.
There are also a great number of R packages only hosted on GitHub. Since
CRAN and Bioconductor packages are not allowed to depend on GitHub
packages, it would be interesting to study how the dependencies
heaviness is transmitted to GitHub packages. Since GitHub packages are
more for experimental purposes, one hypothesis is that GitHub packages
may suffer more from heavy parents.

Evolution of the package ecosystem is also a popular topic in software
engineering \citep{German, Kikas, Mora_evolution} which studies the longitudinal change of packages as well as
their dependency relations over time. In particular, we think the
following two topics might be worth exploring. 1. We can study the
dependency changes after a high impact package was introduced to the
ecosystem. For example, \emph{tidyverse} was introduced to CRAN in 2016.
Since then, it has become a core package for data analysis.
\emph{tidyverse} is a heavy package with HC of 48.4 and it would be
interesting to study how it changes the dependency structure of the ecosystem. 
2. Specific for Bioconductor, it would be interesting
to study the dependency structure change along with the evolution of
high-throughput technologies, e.g., in the era of microarray, genomics,
single cell transcriptomics, and
multi-omics\footnote{An example of the timeline can be found at \url{https://carpentries-incubator.github.io/bioc-project/02-introduction-to-bioconductor/index.html}.}.
This might help to answer the question: does more advanced technology
make the corresponding tools more complex? Finally, we hope the study
presented in this paper as well as the \emph{pkgndep} package can give
developers new insights to properly maintain and optimize dependencies
of their packages, then to build a healthier and more robust R ecosystem
in the future.

In Section \ref{consideration_for_developers}, we manually explored that
heavy dependencies of a group of packages can be traced back to the
upstream package \emph{nloptr} which inherits large dependencies from
\emph{testthat}. This analysis is important because it helps to find out
the ``source'' of the heavy dependency transmission. Then the question is
how to reveal such ``source packages'' systematically and automatically?
To answer this question, we can look at a parent package \emph{A} and
its child \emph{P} from two aspects. First, the dependency transmission
from \emph{A} to \emph{P} is influential in the ecosystem where \emph{A}
contributes a huge amount of dependency heaviness to the downstream via
\emph{P}. Let's denote the total amount of dependency heaviness from
\emph{A} to its downstream via \emph{P} as
\(h_\mathrm{d, total}^{A \rightarrow(P)}\). Using the same denotations in Equation \ref{eq:5},
it can be calculated as

\begin{equation}
  \label{eq:19}
h_\mathrm{d, total}^{A \rightarrow(P)} = \sum_{k=1}^{K_\mathrm{d}}(n_{1k} - n_{2k}) \cdot I(A_k \in S^P_\mathrm{d})
\end{equation}

\noindent where \(A_k\) is \emph{A}'s \textit{k}\textsuperscript{th} downstream package and \(S^P_\mathrm{d}\) is the set of \emph{P}'s downstream packages. Second, heavy dependencies
transmitted to \emph{A}'s downstream via \emph{P} are not originated
from \emph{A}'s parents. This means \emph{A} is the source of the heavy
dependency chain while very few dependencies are accumulated from \emph{A}'s parents. Let's denote \emph{A}'s
MHP parent as \emph{B}, then we can quantitatively measure the level of
\(A \rightarrow P\) being a source of the heaviness transmission denoted as $s^{A \rightarrow(P)}$ by

\begin{equation}
  \label{eq:20}
s^{A \rightarrow(P)} = h_\mathrm{d, total}^{A \rightarrow(P)} - h_\mathrm{d, total}^{B \rightarrow(P)} .
\end{equation}

\noindent In this way, if \emph{A} is the source of heaviness transmission via \emph{P}, $h_\mathrm{d, total}^{B \rightarrow(P)}$ 
will be small which makes $s^{A \rightarrow(P)}$ being a large value; while if most of the heaviness is still from \emph{A}'s parent
\emph{B}, $h_\mathrm{d, total}^{A \rightarrow(P)}$ and $h_\mathrm{d, total}^{B \rightarrow(P)}$ would be similar which makes
$s^{A \rightarrow(P)}$ being a small value.

The dependency heaviness analysis can be extended to ecosystems in other
programming languages. Theoretically, definitions of various heaviness
metrics have no assumption of which package ecosystem to use, and
they can be universally applied as long as a global dependency graph is
available. Similar as \emph{pkgndep}, we plan to implement a general
purpose tool that supports many other package ecosystems also with a
web-based analysis platform for them.

\section{Conclusion}\label{conclusion}

We performed a systematic analysis on the dependency heaviness landscape
of the R package ecosystem. We revealed the general patterns of the
dependency transmission locally from parent to child packages, also
remotely from upstream to downstream packages. Using network analysis
approaches, we revealed top packages and key paths that play significant
roles in transmitting dependencies in the ecosystem. The complete
analysis has been implemented as a web-based database and we believe it
will facilitate researchers as well as R package developers to better
understand the R package ecosystem and to build more robust software.

\section*{CRediT authorship contribution
statement}\label{credit-authorship-contribution-statement}
\addcontentsline{toc}{section}{CRediT authorship contribution statement}

\textbf{Zuguang Gu:} Conceptualization; Formal analysis; Investigation;
Methodology; Software; Visualization; Writing - original draft; Writing
- review \& editing.

\section*{Declaration of competing
interest}\label{declaration-of-competing-interest}
\addcontentsline{toc}{section}{Declaration of competing interest}

The authors declare that they have no known competing financial
interests or personal relationships that could have appeared to
influence the work reported in this paper.

\section*{Data availability}\label{data-availability}
\addcontentsline{toc}{section}{Data availability}

All data as well as the web-based
dependency database are available in the \emph{pkgndep} R package (\url{https://CRAN.R-project.org/package=pkgndep}). How
to access the data is described in Section \ref{data}. The use of the heaviness database is
described in Section \ref{database}. \emph{pkgndep} version 1.2.* can be used to reproduce the analysis in this study.

\section*{Acknowledgements}\label{acknowledgements}
\addcontentsline{toc}{section}{Acknowledgements}

This study did not receive any specific grant from funding agencies in
the public, commercial, or not-for-profit sectors.

\bibliographystyle{elsarticle-harv} 
\bibliography{pkgndep_global_jss_revision}

\end{document}